\documentclass[5p,twocolumn]{elsarticle}

\usepackage{amssymb}

\newcounter{theorem}
\newtheorem{theorem}{Theorem}
\newcounter{lemma}
\newtheorem{lemma}{Lemma}
\newcounter{corollary}
\newtheorem{corollary}{Corollary}
\newcounter{example}
\newtheorem{example}{Example}
\newcounter{proposition}

\begin{document}

\begin{frontmatter}


    
\title{Geometric elements and classification of quadrics in rational B\'ezier form}


\author{A. Cant\'on}
\author{L. Fern\'andez-Jambrina}
\author{E. Rosado Mar\'\i a}
\author{M.J. V\'azquez-Gallo}
\address{Matem\'atica Aplicada\\
Universidad Polit\'ecnica de Madrid\\
E-28040-Madrid, Spain}

\begin{abstract}
In this paper we classify and derive closed formulas for geometric
elements of quadrics in rational B\'ezier triangular form (such as the
center, the conic at infinity, the vertex and the axis of paraboloids
and the principal planes), using just the control vertices and the
weights for the quadric patch.  The results are extended also to
quadric tensor product patches.  Our results rely on using techniques
from projective algebraic geometry to find suitable bilinear forms for
the quadric in a coordinate-free fashion, considering a pencil of
quadrics that are tangent to the given quadric along a conic.  Most of
the information about the quadric is encoded in one coefficient,
involving the weights of the patch, which allows us to tell apart oval from
ruled quadrics.  This coefficient is also relevant to determine the
affine type of the quadric.  Spheres and quadrics of revolution are
characterised within this framework.
\end{abstract}
\begin{keyword} Algebraic projective geometry \sep 
rational B\'ezier patches \sep quadrics \sep Steiner surfaces.
\end{keyword}


\end{frontmatter}

\section{Introduction}

One of the reasons for extending the framework of surfaces used in 
CAD from polynomial or piecewise polynomial surfaces to rational ones 
is the inclusion of quadrics in an exact fashion, since quadric surfaces 
such as spheres, cones, cylinders, paraboloids are commonly used in 
engineering and in architecture \cite{architecture}.

There are mainly two ways of implementing surfaces in CAD: tensor
product patches and B\'ezier triangles \cite{triangle}.  The former
ones are the most common, but the latter have applications in
animation and finite element theory due to the flexibility of
triangles, compared to quadrilaterals, for constructing surfaces and
adapting to different topologies without producing singularities.

Whereas rational quadratic curves are conics, we have a different 
situation when we go to surfaces. On one hand, not every quadric 
triangular patch can be represented as a rational quadratic B\'ezier 
triangle. On the other hand, rational quadratic B\'ezier 
triangles are in general quartic surfaces \cite{sederberg} and only 
in some cases quadric patches are obtained \cite{hansford}. 

Quadric triangular patches have been studied from many points of view.
\cite{lodha} describes a method for constructing quadric patches
grounded on a quadratic B\'ezier control polyhedron.  \cite{dietz} and
\cite{dietz1} use an algebraic approach to construct curves and
surfaces on quadrics.  \cite{coffman} classifies quadratically
parametrised surfaces and studies their geometry.  \cite{degen}
provides a thorough classification of rational quadratic triangles.
\cite{gudrunquadric} provides an algorithm for checking whether a
rational B\'ezier triangle belongs to a quadric and classifying it. 
This solution is shown in \cite{gudrunIEEE} to be one of fifteen 
possibilities, with different numerical conditioning. 
In \cite{gudruninvariants} the algorithm is used to obtain the axes
of the quadric.  \cite{reyes-quadrics} provides a tool for
constructing rational quadratic patches on non-degenerate quadrics.
Three corner points and three weights are used as shape
parameters in \cite{gudrun-parametric} to design quadric surface
patches. In \cite{gudrunmarco} the shape parameters are three points 
and the normal vectors at them.

In this paper we draw geometric information from rational B\'ezier
quadric triangles and tensor product patches.  We classify them and
calculate their geometric elements in closed form, using just the
control net and the weights of the patch, as it is done for
conics in \cite{conics}.  One of the reasons for
doing this is that closed formulas for geometric elements of a quadric
can be of great help for designing with them.

With this goal in mind, we derive bilinear forms for the quadrics,
both in point and tangential form in a coordinate-free fashion, using
techniques borrowed from algebraic projective geometry, such as the
use of pencils of quadrics, which have also been used in 
\cite{gudrun-parametric,gudrunmarco,gudruntesis}.  Using linear forms with clear geometric
meaning (tangent planes, planes containing boundary conics\ldots)
instead of, for instance, cartesian coordinates, allows us to find a
closed form for the implicit equation of these quadrics in terms of
their weights, their tangent planes and the plane spanned by the
corner vertices of the patch.  The major originality and
advantage of our formulation is the use  of only geometric information
about the patch, avoiding the use of coordinate expressions as an
intermediate step.  In fact, most of the relevant geometric
information of the surface which is necessary for its classification
is comprised in a single coefficient, the parameter of the pencil of
quadrics, which depends on the weights of the patch.

The coordinate-free bilinear forms for the quadrics enable the 
derivation of closed formulas for several
geometric elements (center of the quadrics, bilinear forms for the 
conic at infinity, diametral planes) calculated in terms of
weights and vertices. To our knowledge, such closed formulas have not been 
produced before. Similarly, linear forms for principal planes are obtained 
up to solving a cubic equation. The degeneracy of the solutions of 
this equation allows us to identify  quadrics of revolution and 
spheres.

Finally, the previous results for triangular quadric patches can be 
extended to rational biquadratic quadric patches using the tangent 
planes at three corners of the boundary of the tensor product patch.

This paper is organised as follows: We revisit rational B\'ezier
triangular patches and the characterisation of quadric patches in
Section~2.  In Section~3 we construct the pencil of quadrics, in point
and in tangential form, which are tangent to our quadric at the conic
through the three corner vertices of the patch and fix the only free
parameter in terms of the weights of the quadric.  This provides us
bilinear forms for the quadrics and hence implicit equations.  A
coordinate-free expression for the center of the quadric is obtained
in Section~4 as a barycentric combination of the corner vertices of
the patch and the intersection of their tangent planes.  It is used to
tell paraboloids from centered quadrics.  In Section~5 we calculate a
bilinear form for the conic at infinity, which is useful for
determining the affine type of the quadric.  Section~6 is devoted to
degenerate quadrics.  With this information we provide in Section~7 a
way to classify quadric patches using the signature of the bilinear
form, the center and the bounding conic arcs.  We introduce the scalar
product in Section~8 in order to calculate Euclidean geometric
elements such as principal planes and axes of quadrics and vertices of
paraboloids.  This allows us to characterise spheres and quadrics of
revolution.  In Section~9 we show several examples of application of our
results. We show how to extend our results to quadric tensor product 
patches in Section~10. A final section of conclusions is included.

\section{Quadratic rational B\'ezier triangular patches}

Rational B\'ezier triangular patches of degree
$n$ are defined using trivariate Bernstein polynomials of degree $n$ in parameters 
$u,v,w$,
\[B^{n}_{ijk}(u,v,w)=\frac{n!}{i!j!k!}u^iv^jw^k,\quad i+j+k=n,\]
\begin{equation}\label{steiner}c(u,v,w)=
\frac{\displaystyle\sum_{i+j+k=n}\omega_{ijk}c_{ijk}B^n_{ijk}(u,v,w)}
{\displaystyle\sum_{i+j+k=n}\omega_{ijk}B^n_{ijk}(u,v,w)},
\end{equation} such that $u+v+w=1$, $u,v,w\in[0,1]$, where the
coefficients $c_{ijk}$ are the control points for the surface and
$\omega_{ijk}$ are their corrresponding weights. 

The surface patch is bounded by three rational B\'ezier curves of
degree $n$, which are obtained by fixing $u=0$, $v=0$, $w=0$.  Their
respective control points and weights are respectively the ones with
index $i=0$, $j=0$, $k=0$.

We are interested in the special case of quadratic surfaces. In this 
case the control net and the matrix of weights are
\begin{equation}
\left[\begin{array}{c}c_{002}\hspace{0.2cm}c_{011}\hspace{0.2cm}c_{020}\\
c_{101}\hspace{0.2cm}c_{110}\\ 
c_{200}\end{array}\right],\qquad
\left[\begin{array}{c}\omega_{002}\hspace{0.2cm}\omega_{011}\hspace{0.2cm}\omega_{020}\\
\omega_{101}\hspace{0.2cm}\omega_{110}\\ \omega_{200}\end{array}\right],
\end{equation}
and the surface patch is completely determined by the three boundary 
conic curves:

The conic at $u=0$ has control points 
$\{c_{002},c_{011},c_{020}\}$ and weights  
$\{\omega_{002},\omega_{011},\omega_{020}\}$, the one at $v=0$ 
has control points $\{c_{002}, c_{101}, c_{200}\}$ and weights $\{\omega_{002},\omega_{101},\omega_{200}\}$, 
and the one at $w=0$ has control points
$\{c_{020},c_{110},c_{200}\}$ and weights 
$\{\omega_{020},\omega_{110},\omega_{200}\}$. We just consider the 
case of non-degenerate boundary curves. We name the planes where such 
conics are located respectively as $u,v,w$.

Quadratic rational B\'ezier triangular surfaces comprise quadrics as 
a subcase, but in general they are quartic surfaces named Steiner 
surfaces \cite{sederberg}. On the contrary, not every quadric patch 
bounded by three arcs can be parametrised as a quadratic rational 
B\'ezier triangle. According to \cite{hansford}: 
\begin{itemize}
\item If the Steiner surface is a non-degenerate quadric, the
three conic boundary curves meet at a point $S$ and their
respective tangent vectors at $S$ define a plane.

\item If the three conic boundary curves meet at a point $S$ and their
respective tangent vectors at $S$ define a plane, the Steiner surface is a
quadric patch.

\item If there are three alligned points, one on each boundary conic
curve, and the tangent plane to the surface is the same for all three 
points, the Steiner surface is a degenerate quadric patch.
\end{itemize}

The last case just implies that the quadric (cylinder or cone) is
ruled and the three points lie on the same ruling.

We are considering Steiner surfaces with such a point $S$. This case 
comprises non-degenerate quadrics. Since $S$ belongs to the 
three boundary conics, there are values $u_{s}, v_{s}, w_{s}$ of the 
parameters such that
\begin{eqnarray*}\hspace{-0.5cm}S&=&\frac{\displaystyle\sum_{j=0}^2
\omega_{0j2-j}c_{0j2-j}B^{2}_{j}(v_{s})}
{\displaystyle\sum_{j=0}^2
\omega_{0j2-j}B^{2}_{j}(v_{s})}=\frac{\displaystyle\sum_{i=0}^2
\omega_{i2-i0}c_{i2-i0}B^{2}_{i}(u_{s})}
{\displaystyle\sum_{i=0}^2
\omega_{i2-i0}B^{2}_{i}(u_{s})}\\\hspace{-0.5cm}&=&\frac{\displaystyle\sum_{k=0}^2
\omega_{2-k0k}c_{2-k0k}B^{2}_{k}(w_{s})}
{\displaystyle\sum_{k=0}^2
\omega_{2-k0k}B^{2}_{k}(w_{s})},\end{eqnarray*}
where we have introduced quadratic Bernstein polynomials,
\[B^{2}_{0}(t)=(1-t)^2,\quad B^2_{1}(t)=2t(1-t),\quad 
B^2_{2}(t)=t^2.\]

However, since the set of weights of rational B\'ezier curves is
unique up to M\"obius transformations \cite{farin} of the parameter, 
we can use this freedom to set $u_{s}$, $v_{s}$, $w_{s}$ at infinity. 
It is a simple exercise to check that this is possible if and only if 
the tangent vectors at $S$ are coplanar \cite{oldquadric}. Hence, this choice is not a 
restriction. 

From now on we assume that the set of weights for the surface 
fulfills this condition and therefore we may write $S$ in three 
different forms as
\begin{eqnarray}\label{baryS}S&=&
\frac{\omega_{002}c_{002}-2\omega_{011}c_{011}+ \omega_{020}c_{020}}{\omega_{u}}
\nonumber\\&=&
\frac{\omega_{002}c_{002}-2\omega_{101}c_{101}+\omega_{200}c_{200}}{\omega_{v}}
\nonumber\\&=&
\frac{\omega_{200}c_{200}-2\omega_{110}c_{110}+\omega_{020}c_{020}}{\omega_{w}},
\end{eqnarray}
where $\omega_{u},\omega_{v},\omega_{w}$ guarantee that these are 
barycentric combinations,
\begin{eqnarray*}\label{denoms}\hspace{-0.5cm}
&\omega_{u}=\omega_{002}-2\omega_{011}+ \omega_{020},\ 
\omega_{v}=\omega_{002}-2\omega_{101}+ \omega_{200},&\\ \hspace{-0.5cm}
&\omega_{w}=\omega_{200}-2\omega_{110}+\omega_{020},&
\end{eqnarray*}
if $S$ is a proper point. If $S$ is point at infinity, the 
coefficients $\omega_{u},\omega_{v},\omega_{w}$ are chosen in order 
to have the same vector as representantive for $S$ on the three 
boundary curves.

%


\section{Pencils of Steiner quadrics}

In order to obtain a bilinear form for the quadric, we describe the
pencil of quadrics which are tangent to our quadric at the conic on
the plane $t$ defined by the points at the corners of the surface
patch, $P=c_{002}$, $Q=c_{020}$, $R=c_{200}$.  This pencil of quadrics
has been used also in
\cite{gudrun-parametric,gudrunmarco,gudruntesis}.

To this aim, we just need two quadrics belonging to the pencil
\cite{kneebone}, which can be degenerate.
\begin{figure}
\centering
\includegraphics[height=3cm]{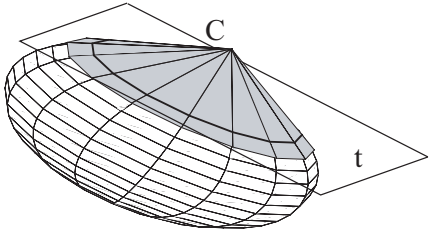}
\caption{Tangent cone to a quadric along a conic on a plane $t$}
\label{tangentcon}
\end{figure}

For instance, we can take the cone $C$ tangent to 
our quadric along the conic on $t$ and the double plane $t$ (see 
Fig.~\ref{tangentcon}). Hence, 
the bilinear form for this pencil of quadrics is just $
C +\lambda t^2$, using the same letter for a surface and its form. Finally, we can determine our quadric just using the 
existence of point $S$,
\[ C(S)+\lambda t(S)^{2}=0.\]

We just have then to provide a bilinear form for the cone $C$. Since 
$\lambda=-C(S)/t(S)^{2}$, we notice that the sign of $\lambda$ 
depends on whether $S$ lies in or out of the cone $C$.

\begin{figure}
\centering
\includegraphics[height=4cm]{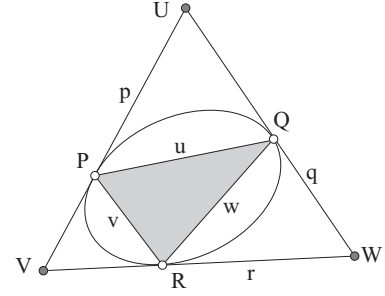}
\caption{Conic circumscribed by a triangle}
\label{circumscribe}
\end{figure}

It will be useful to start with the bilinear form for the conic on $t$
(see Fig.~\ref{circumscribe}).  In this figure $u,v,w$ are the
straight lines which are respectively the intersections of the planes
$u,v,w$ (where the boundary conics lie) with the plane $t$.

\begin{figure}
\centering
\includegraphics[height=4cm]{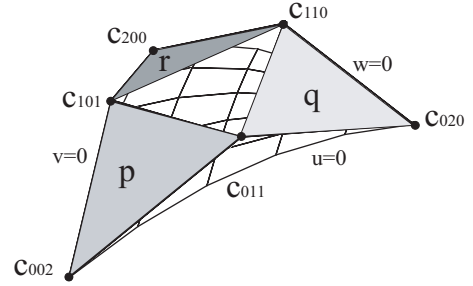}
\caption{Tangent planes $p,q,r$}
\label{plantan}
\end{figure}

We use the same letter for a point (uppercase) and its polar line 
(lower case).  If a point $A$ lies on the conic, its polar line is the 
tangent $a$ to the conic at such point. Hence, $p$, $q$, $r$ are 
respectively the tangent lines to the conic at $P$, $Q$, $R$ (see 
Fig.~\ref{plantan}). We 
shall use the same letters $p,q,r$ also for the tangent planes at $P,Q,R$.

On the other hand, the polar line to a point $A$ not on the conic is the 
line $a$ linking the tangency points of the tangent lines to the 
conic drawn from $A$. Hence, $u$, $v$, $w$ are the polar lines to the 
vertices of the circumscribed triangle $U$, $V$, $W$.

The pencil of conics circumscribed by the triangle defined by the 
lines $p$, $q$, 
$r$ is easily described in tangential form. The bilinear form for 
this pencil is just
\[\alpha UV+\beta UW+\gamma VW,\] with coefficients $\alpha$, 
$\beta$, $\gamma$. It is easy to check that this form vanishes on 
$p$, $q$, $r$,
since \[
\hspace{-5mm}p(U)=0=p(V),\ q(U)=0=q(W),\ r(V)=0=r(W).\]

But, for our purposes, we require the point bilinear form for the 
conic. Referred to the lines $r$, $q$, $p$, which form a dual reference to the one 
formed by $U$, $V$, $W$, provided that their 
linear forms satisfy 
\begin{equation}\label{dual}p(W)=1,\quad q(V)=1, \quad r(U)=1,\end{equation}
the matrix of this point bilinear form is the inverse of the one for the 
tangential bilinear form,
\[\hspace{-5mm}\left(
\begin{array}{ccc} 
0 & \alpha & \beta \\ \alpha & 0 & \gamma \\ \beta & \gamma & 0 
\end{array}
\right)^{-1} =-\frac{1}{2\alpha\beta\gamma}
 \left(
\begin{array}{ccc} 
\gamma^2 & -\beta\gamma & -\alpha\gamma \\ \
-\beta\gamma & \beta^2 & -\alpha\beta \\ 
-\alpha\gamma & -\alpha\beta & \alpha^2 
\end{array}
\right),\]
and hence such a bilinear form is
\[\alpha^2p^{2}-2\alpha\beta 
pq-2\alpha\gamma pr+\beta^2q^2-2\beta\gamma qr+\gamma^2 r^2.\]

In order to determine the coefficients $\alpha, \beta,\gamma$, it 
is useful to take into account that (\ref{baryS}) provides relations 
between values of the linear forms for $p,q,r$,
\begin{eqnarray*}
\frac{\omega_{020}}{\omega_{u}}p(Q)=p(S)=\frac{\omega_{200}}{\omega_{v}}p(R),\\
\frac{\omega_{200}}{\omega_{w}}q(R)=q(S)=\frac{\omega_{002}}{\omega_{u}}q(P),\\
\frac{\omega_{002}}{\omega_{v}}r(P)=r(S)=\frac{\omega_{020}}{\omega_{w}}r(Q),
\end{eqnarray*}
which can be used to write $U,V,W$ in terms of $P,Q,R$,
\begin{eqnarray}\label{UVW}
U=\frac{\omega_{002}\omega_{w}P+\omega_{020}\omega_{v}Q-\omega_{200}\omega_{u}R}
{\Omega_{U}},\nonumber\\
V=\frac{\omega_{002}\omega_{w}P-\omega_{020}\omega_{v}Q+\omega_{200}\omega_{u}R}
{\Omega_{V}},\nonumber\\
W=\frac{-\omega_{002}\omega_{w}P+\omega_{020}\omega_{v}Q+\omega_{200}\omega_{u}R}
{\Omega_{W}},
\end{eqnarray}
or, conversely,
\[\hspace{-7mm}
P=\frac{\Omega_{U}U+\Omega_{V}V}{2\omega_{002}\omega_{w}},
Q=\frac{\Omega_{W}W+\Omega_{U}U}{2\omega_{020}\omega_{v}},
R=\frac{\Omega_{V}V+\Omega_{W}W}{2\omega_{200}\omega_{u}},\]
where the normalisation denominators are
\begin{eqnarray}\label{Omegas}
\Omega_{U}=\omega_{002}\omega_{w}+\omega_{020}\omega_{v}-\omega_{200}\omega_{u},
\nonumber\\
\Omega_{V}=\omega_{002}\omega_{w}-\omega_{020}\omega_{v}+\omega_{200}\omega_{u},
\nonumber\\
\Omega_{W}=-\omega_{002}\omega_{w}+\omega_{020}\omega_{v}+\omega_{200}\omega_{u},
\end{eqnarray}
provided that $U,V,W$ are proper points. If one of these points goes 
to infinity, the respective denominator is taken to be one. The 
interpretation of these coefficients as weights for $U,V,W$ on the 
conic on $t$ is discussed in Section~6.

Finally, the normalisation condition (\ref{dual}) allows us to fix 
the linear forms for $p,q,r$ since
\begin{eqnarray}\label{SS}&
p(W)=\displaystyle\frac{2\omega_{u}\omega_{v}}{\Omega_{W}}p(S),\qquad 
q(V)=\frac{2\omega_{u}\omega_{w}}{\Omega_{V}}q(S),&\nonumber\\
&r(U)=\displaystyle\frac{2\omega_{v}\omega_{w}}{\Omega_{U}}r(S).&
\end{eqnarray}

Since we require that points $P$, $Q$, $R$ lie on the conic,
\[ \beta q(P)=\gamma r(P),\quad
\alpha p(Q)=\gamma r(Q),\quad
\alpha p(R)=\beta q(R),\]
we identify the coefficients of the bilinear form,
\[ \alpha=\frac{1}{\Omega_{W}},\quad \beta=\frac{1}{\Omega_{V}},\quad 
\gamma=\frac{1}{\Omega_{U}},\]
up to a common factor. 

Hence, a bilinear form for the conic on $t$ is
\begin{equation}\label{tconic}
\frac{p^2}{\Omega_{W}^2}+\frac{q^2}{\Omega_{V}^2}+\frac{r^2}{\Omega_{U}^2}
-\frac{2pq}{\Omega_{W}\Omega_{V}}-\frac{2pr}{\Omega_{W}\Omega_{U}}
-\frac{2qr}{\Omega_{V}\Omega_{U}}
\end{equation}
and, in tangential form,
\begin{equation}
\Omega_{U}\Omega_{V}UV+\Omega_{U}\Omega_{W}UW+\Omega_{V}\Omega_{W}VW.
\end{equation}

The terms of the bilinear form can be factored as
\[\hspace{-5mm}
\left(\frac{p}{\Omega_{W}}-\frac{q}{\Omega_{V}}-\frac{r}{\Omega_{U}}\right)^2
+\left(\frac{q}{\Omega_{V}}-\frac{r}{\Omega_{U}}\right)^2-
\left(\frac{q}{\Omega_{V}}+\frac{r}{\Omega_{U}}\right)^2,\]
and thereby it has signature $(++-)$ regardless of the 
values of the coefficients.



We move now back from plane to space: If $p$, $q$, $r$ are the tangent planes to the quadric at
$P$, $Q$, $R$, the bilinear form (\ref{tconic}) describes a \emph{degenerate}
quadric of signature $(++-\,0)$ which contains the conic on $t$  and it is tangent to our 
quadric along it. Since the rank of the bilinear form is three, it is a cone (or 
a cylinder, if the intersection point of $p$, $q$, $r$ goes to 
infinity), since these are the only non-plane degenerate quadrics. 
Hence, it is the cone we are looking for.

The bilinear form for the pencil of quadrics referred to the planes 
$r,q,p,t$,
\[
\frac{p^2}{\Omega_{W}^2}+\frac{q^2}{\Omega_{V}^2}+\frac{r^2}{\Omega_{U}^2}
-\frac{2pq}{\Omega_{W}\Omega_{V}}-\frac{2pr}{\Omega_{W}\Omega_{U}}
-\frac{2qr}{\Omega_{V}\Omega_{U}}+\lambda t^2,\]
produces a bilinear form for our quadric if
$\lambda=-C(S)/t(S)^{2}$. 

We see that the sign of $\lambda$ determines the signature of the 
bilinear form, defined as the difference between the number of its positive 
and negative eigenvalues. This is relevant since ruled quadrics have 
null signature, $(++--)$ and oval quadrics have signature two $(++-+)$. 
This is useful to classify the quadric.

In tangential form referred to the points $U,V,W,T$, the bilinear 
form for the quadric is
\[
\Omega_{U}\Omega_{V}UV+\Omega_{U}\Omega_{W}UW+\Omega_{V}\Omega_{W}VW
-\frac{T^2}{\lambda},
\]
if $t(T)=1$, where $T$ is the intersection point of the planes 
$p,q,r$, that is, the vertex of the cone $C$.

With this information we can write $T$ as
\begin{eqnarray}\hspace{-0.5cm}\label{TTT}
T\!\!\!&=&\!\!\!\frac{\omega_{002}\omega_{w}(\omega_{u}+\omega_{v}-\omega_{w})P+
\omega_{020}\omega_{v}(\omega_{u}-\omega_{v}+\omega_{w})Q}
{\epsilon_{U}\Omega_{U}\omega_{u}+\epsilon_{V}\Omega_{V}\omega_{v}+
\epsilon_{W}\Omega_{W}\omega_{w}-2\epsilon_{S}\omega_{u}\omega_{v}\omega_{w}}
\nonumber\\ 
\hspace{-0.5cm}
&+&\frac{
\omega_{200}\omega_{u}(-\omega_{u}+\omega_{v}+\omega_{w})R-
2\omega_{u}\omega_{v}\omega_{w}S}
{\epsilon_{U}\Omega_{U}\omega_{u}+\epsilon_{V}\Omega_{V}\omega_{v}+
\epsilon_{W}\Omega_{W}\omega_{w}-2\epsilon_{S}\omega_{u}\omega_{v}\omega_{w}},
\end{eqnarray}
or alternatively,
\begin{eqnarray*}\hspace{-0.5cm}
S\!\!\!&=&\!\!\!\frac{\Omega_{U}}{2\omega_{v}\omega_{w}}U+
\frac{\Omega_{V}}{2\omega_{u}\omega_{w}}V+
\frac{\Omega_{W}}{2\omega_{u}\omega_{v}}W+\\ 
\hspace{-0.5cm}
&+&\frac{2\epsilon_{S}\omega_{u}\omega_{v}\omega_{w}-
\epsilon_{U}\Omega_{U}\omega_{u}-\epsilon_{V}\Omega_{V}\omega_{v}-
\epsilon_{W}\Omega_{W}\omega_{w}}
{2\omega_{u}\omega_{v}\omega_{w}}T,
\end{eqnarray*}
where $\epsilon_A$ takes the value one if the 
corresponding point $A$ is proper or zero if it is a point at infinity.

Now we can compute $\lambda$ and determine the quadric:

\begin{theorem}\label{bilin}A rational triangular quadratic patch with control
net $\{c_{002}, 
c_{011},c_{020},c_{101},c_{110},c_{200}\}$ and weights 
$\{\omega_{002}, \omega_{011},\omega_{020},\omega_{101},\omega_{110},\omega_{200}
\}$, such that the three boundary conics meet at a point $S$, which is written as in 
(\ref{baryS}), is a quadric with bilinear form
\begin{equation}\label{bilform}\hspace{-7mm}\mathcal{Q}=
\frac{p^2}{\Omega_{W}^2}+\frac{q^2}{\Omega_{V}^2}+\frac{r^2}{\Omega_{U}^2}
-\frac{2pq}{\Omega_{W}\Omega_{V}}-\frac{2pr}{\Omega_{W}\Omega_{U}}
-\frac{2qr}{\Omega_{V}\Omega_{U}}+\lambda t^2,\ 
\end{equation}
where $t$ is the linear form of the plane
containing $c_{002},c_{020}, c_{200}$, $p$ is the linear form of the  plane containing
$c_{002}$, $c_{011}$, $c_{101}$, $q$ is the linear form of the plane
containing $c_{020}$, $c_{011}$, $c_{110}$ and $r$ is the linear form
of the plane containing $c_{200}$, $c_{101}$, $c_{110}$ which
satisfy
\[p(W)=1, \quad q(V)=1, \quad r(U)=1,\quad t(T)=1,\]
at the points $U$ (intersection of the planes $p,q,t$), $V$
 (intersection of the planes $p,r,t$) and $W$ (intersection of the 
planes $q,r,t$) given by (\ref{UVW}) and the point $T$ (intersection of the 
planes $p,q,r$), given by  (\ref{TTT}).

The coefficients are given in (\ref{Omegas}) except for  $\lambda$, 
\begin{equation}\label{lamda}\hspace{-7mm}
\lambda=-\frac{\omega_{u}^2+\omega_{v}^2+\omega_{w}^2-2\omega_{u}\omega_{v}-
2\omega_{v}\omega_{w}-2\omega_{w}\omega_{u}}
{(2\epsilon_{S}\omega_{u}\omega_{v}\omega_{w}-
\epsilon_{U}\Omega_{U}\omega_{u}-\epsilon_{V}\Omega_{V}\omega_{v}-
\epsilon_{W}\Omega_{W}\omega_{w})^2},
\end{equation}
and the tangential bilinear form for the quadric is
\begin{equation}\label{tangential}\tilde\mathcal{Q}=
\Omega_{U}\Omega_{V}UV+\Omega_{U}\Omega_{W}UW+\Omega_{V}\Omega_{W}VW
-\frac{T^2}{\lambda},
\end{equation}

Furthermore, for $\lambda>0$ the quadric is oval and for $\lambda<0$ 
the quadric is ruled.\end{theorem}

The expression we have obtained for the bilinear form of the quadric has the 
advantage of encoding most of the information about the surface in just the 
coefficient $\lambda$ for $t^2$. 

This result provides a procedure for computing a bilinear form, 
and hence the implicit equation, for a non-degenerate Steiner quadric 
patch in a coordinate-free fashion using just the vertices of the 
control net and their respective weights:

\begin{enumerate}
    \item Obtain $S$ as intersection of the planes $u,v,w$.

    \item  Compute the normalised linear forms for the planes $t,p,q,r$.

    \item  Obtain an equivalent list of weights fulfilling 
    (\ref{baryS}).

    \item Use Theorem~\ref{bilin} to obtain the bilinear form
    $\mathcal{Q}$ for the quadric patch.
    
    \item  The implicit equation for the quadric patch is then 
    $\mathcal{Q}(X,X)=0$.
\end{enumerate}

\section{Center of a quadric}

The bilinear form for the quadric provides a way to obtain its center 
$Z$ as the pole of the plane $z$ at infinity. Since the plane at 
infinity is formed by vectors, we may write its elements as 
barycentric combinations
\[a U+bV+cW+dT,\qquad a+b+c+d=0.\]

We have to consider the possibility of having any of the points of the
reference at infinity. In such case, the null sum is restricted to 
proper points,
\begin{equation} 
a U+bV+cW+dT,\quad 
\epsilon_{U}a+\epsilon_{V}b+\epsilon_{W}c+\epsilon_{T}d=0.\end{equation}

Hence, a linear form for the plane at infinity in this reference is 
just
\begin{equation}
z=\epsilon_{W}p+\epsilon_{V}q+\epsilon_{U}r+\epsilon_{T}t.
\end{equation}

The pole of this plane,
\begin{eqnarray}\hspace{-5mm}\label{centerpole}
\tilde\mathcal{Q}(z)
&=&
\epsilon_{W}\Omega_{W}(\Omega_{U}U+\Omega_{V}V)+
\epsilon_{V}\Omega_{V}(\Omega_{W}W+\Omega_{U}U)\nonumber\\ 
\hspace{-5mm}
&+&\epsilon_{U}\Omega_{U}(\Omega_{V}V+\Omega_{W}W)-\frac{2\epsilon_{T}}{\lambda}
T,
\end{eqnarray}
can be written in a simpler way in terms of $P,Q,R,T$ in order to 
produce an expression for the center of the quadric,
\begin{eqnarray}\label{center}\hspace{-7mm}
Z&=&\frac{\epsilon_{W}\Omega_{W}\omega_{w}\omega_{002}P+
\epsilon_{V}\Omega_{V}\omega_{v}\omega_{020}Q+
\epsilon_{U}\Omega_{U}\omega_{u}\omega_{200}R}{\Omega_{Z}}\nonumber\\\hspace{-7mm}
&-&\frac{\epsilon_{T}}{\lambda\Omega_{Z}}T,\ 
\end{eqnarray}
where the denominator $\Omega_Z$ is one if the center is a point at 
infinity or, if it is a proper point,
\begin{equation}\label{paraboloid}
\Omega_{Z}=\tilde\Omega_{W}\omega_{w}\omega_{002}+
\tilde\Omega_{V}\omega_{v}\omega_{020}+
\tilde\Omega_{U}\omega_{u}\omega_{200}-\frac{\epsilon_{T}}{\lambda},
\ \end{equation}
and we have introduced for simplicity,
\begin{equation}\label{tildeOmegas} \tilde\Omega_{U}=\epsilon_{U}\Omega_{U},\quad 
\tilde\Omega_{V}=\epsilon_{V}\Omega_{V},\quad
\tilde\Omega_{W}=\epsilon_{W}\Omega_{W}.\end{equation}

Since the center of a paraboloid is a point at infinity, we have a 
simple characterisation:
\begin{corollary}A rational triangular quadratic patch with control
net $\{c_{002}, 
c_{011},c_{020},c_{101},c_{110},c_{200}\}$ and weights 
$\{\omega_{002}, \omega_{011},\omega_{020},\omega_{101},\omega_{110},\omega_{200}
\}$, such that the three boundary conics meet at a point $S$, which is written as in 
(\ref{baryS}), is a paraboloid if the quadric is non-degenerate and
\[\hspace{-5mm}
\tilde\Omega_{W}\omega_{w}\omega_{002}+
\tilde\Omega_{V}\omega_{v}\omega_{020}+
\tilde\Omega_{U}\omega_{u}\omega_{200}-\frac{\epsilon_{T}}{\lambda}=0, \]
where the coefficients are given in (\ref{Omegas}), (\ref{lamda}) and 
(\ref{tildeOmegas}). 
\end{corollary}

\section{The conic at infinity}

The conic at infinity is the intersection of the quadric with the 
plane $z$ at infinity and it is formed by its asymptotic directions. 
It is useful for classifying quadric patches, as it is done in 
\cite{gudrunquadric}. 
Since points $X$ on $z$ satisfy
\[z(X)=\epsilon_{W}p(X)+\epsilon_{V}q(X)+\epsilon_{U}r(X)+\epsilon_{T}t(X)=0,\]
we can use as bilinear form for the conic at infinity on $z$
\begin{eqnarray}\label{infconic}
\hspace{-5mm}\mathcal Z&=&\frac{p^2}{\Omega_{W}^2}+\frac{q^2}{\Omega_{V}^2}+\frac{r^2}{\Omega_{U}^2}
-\frac{2pq}{\Omega_{W}\Omega_{V}}-\frac{2pr}{\Omega_{W}\Omega_{U}}
-\frac{2qr}{\Omega_{V}\Omega_{U}}\nonumber\\\hspace{-5mm}&+&\lambda 
(\epsilon_{W}p+\epsilon_{V}q+\epsilon_{U}r)^2,
\end{eqnarray}
except  when $T$ is a point at infinity.

In order to draw information about the conic at infinity, we may 
factor its bilinear form,
\begin{eqnarray*}
\mathcal 
Z&=&(\tilde\Omega_{W}^2\lambda+1)\sigma_{1}^2+
\frac{\lambda}{\tilde\Omega_{W}^2\lambda+1}\sigma_{2}^2\nonumber\\&+&
\left(\tilde\Omega_{U}\tilde\Omega_{V}+\tilde\Omega_{V}\tilde\Omega_{W}+
\tilde\Omega_{W}\tilde\Omega_{U}-
\frac{1}{\lambda}\right)
\sigma_{3}^2,
\end{eqnarray*}
where we have introduced three linear forms, $\sigma_{1}, \sigma_{2}, 
\sigma_{3}$, in order to diagonalise 
$\mathcal Z$, 
\[\hspace{-7mm}
\sigma_{1}:=\frac{p}{\Omega_{W}}+\frac{\tilde
\Omega_{V}\tilde\Omega_{W}\lambda-1}{\tilde\Omega_{W}^2\lambda+1}
\frac{q}{\Omega_{V}}+\frac{\tilde\Omega_{U}\tilde\Omega_{W}\lambda-1}
{\tilde\Omega_{W}^2\lambda+1}
\frac{r}{\Omega_{U}},
\]
\begin{eqnarray*}\sigma_{2}\!\!\!&:=&\!\!\!
(\tilde\Omega_{V}+\tilde\Omega_{W})\frac{q}{\Omega_{V}}
\\\!\!\!&+&\!\!\!\frac{\tilde\Omega_{U}\tilde\Omega_{V}+\tilde\Omega_{V}\tilde\Omega_{W}+
\tilde\Omega_{W}\tilde\Omega_{U}-\tilde\Omega_{W}^2-2/\lambda}
{\tilde\Omega_{V}+\tilde\Omega_{W}}\frac{r}{\Omega_{U}},
\end{eqnarray*}
\[\sigma_{3}:=\frac{2}{\tilde\Omega_V+\tilde\Omega_{W}}\frac{r}{\Omega_{U}}.\]

The case of $T$ at infinity is simpler, as points at infinity satisfy
\[z(X)=\epsilon_{W}p(X)+\epsilon_{V}q(X)+\epsilon_{U}r(X)=0,\]
but it can be handled similarly. 

Combining both cases, we obtain a general expression for the
determinant of $\mathcal Z$ in this reference,
\begin{equation}\label{determinant}
\det \mathcal Z=\lambda\left(\tilde\Omega_{U}\tilde\Omega_{V}+
\tilde\Omega_{V}\tilde\Omega_{W}+\tilde\Omega_{W}\tilde\Omega_{U}
\right)-\epsilon_{T}.\end{equation} 

The conic at infinity of a paraboloid is degenerate.  Hence,
$\det\mathcal Z$ vanishes for these quadrics.  This condition is
equivalent to the one obtained in the previous section.


The classification of the conic at infinity allows us to finish the 
classification of quadrics. Since paraboloids are non-centered, 
one-sheeted hyperboloids are ruled and centered, we just have to tell 
ellipsoids from two-sheeted hyperboloids, since they are both oval 
and centered:


Since $\lambda$ is positive in this case, we notice that if the 
determinant (\ref{determinant}) is positive, the signature of the 
bilinear form is $(+++)$ and hence the conic is imaginary. We have an 
ellipsoid then, since it does not intersect the plane at infinity.

On the other hand, if the 
determinant (\ref{determinant}) is negative, the signature of the 
bilinear form is $(++-)$ and we have a proper conic. The quadric is 
a hyperboloid in this case:

\begin{corollary}A rational triangular quadratic patch with control
net $\{c_{002}, 
c_{011},c_{020},c_{101},c_{110},c_{200}\}$ and weights 
$\{\omega_{002}, \omega_{011},\omega_{020},\omega_{101},\omega_{110},\omega_{200}
\}$, such that the three boundary conics meet at a point $S$, which is written as in 
(\ref{baryS}) and with positive $\lambda$, given by (\ref{lamda}), is an oval quadric and:
\begin{itemize}
\item  An ellipsoid if $\det\mathcal Z>0$.

\item  An elliptic paraboloid if $\det\mathcal Z=0$.

\item  A two-sheeted hyperboloid if $\det\mathcal Z<0$.
\end{itemize}
where the expression for $\det\mathcal Z$ is given in (\ref{determinant}).
\end{corollary}

\section{Conics on planes $t,u,v,w$ and degenerate quadrics}

In order to draw more information about the surface patch, we take a 
look at the conic arcs on planes $t,u,v,w$.

A conic with weights $\omega_{0},\omega_{1},\omega_{2}$  can be 
classified \cite{farin} using the canonical weight 
$\omega=\omega_{1}/\sqrt{\omega_{0}\omega_{2}}$: 

If $\omega<1$, it is an 
ellipse; if $\omega=1$, it is a parabola and if $\omega>1$, it is a 
hyperbola.

The canonical weights for the conics on planes $u,v,w$ are 
respectively
\[\frac{\omega_{011}}{\sqrt{\omega_{002}\omega_{020}}},\qquad
\frac{\omega_{101}}{\sqrt{\omega_{002}\omega_{200}}},\qquad
\frac{\omega_{110}}{\sqrt{\omega_{020}\omega_{200}}}.\]

A set of weights for an arc of the conic on $t$ is readily obtained.
For instance, we can use $P,U,Q$ as control polygon and use the same
kind of construction of ($\ref{baryS}$) to assign an infinite parameter 
to the point $R$, so that its coordinates referred to $P,U,Q$,
\[R=\frac{\omega_{002}\omega_{w}P-\Omega_{U}U+\omega_{020}\omega_{v}Q}{\omega_{200}\omega_{u}},
\]
provides us a set of weights $\omega_{002}\omega_{w}$, 
$\Omega_{U}/2$, $\omega_{020}\omega_{v}$ for the conic arc. 

Similarly, we obtain weights $\omega_{002}\omega_{w}$, $\Omega_{V}/2$,
$\omega_{200}\omega_{u}$ for the arc with control polygon $P,V,R$ and
$\omega_{020}\omega_{v}$, $\Omega_{W}/2$, $\omega_{200}\omega_{u}$ for
the arc with control polygon $Q,W,R$. Hence the canonical weights for 
these arcs are 
\begin{eqnarray}\hspace{-10mm}\label{canonical}
&\omega_{PUQ}=\displaystyle\frac{\Omega_{U}}
{2\sqrt{\omega_{002}\omega_{020}\omega_{v}\omega_{w}}},\quad
\omega_{PVR}=\frac{\Omega_{V}}
{2\sqrt{\omega_{002}\omega_{200}\omega_{u}\omega_{w}}},&\nonumber\\\hspace{-10mm}
&\omega_{QWR}=\displaystyle\frac{\Omega_{W}}
{2\sqrt{\omega_{020}\omega_{200}\omega_{u}\omega_{v}}}.&\end{eqnarray}

We can use either of these to classify the conic on $t$.

Furthermore, this result furnishes an interpretation of $\Omega_{U}$, 
$\Omega_{V}$, $\Omega_{W}$ as weights for the points $U,V,W$ if 
$\omega_{002}\omega_{w}$, $\omega_{020}\omega_{v}$ and 
$\omega_{200}\omega_{u}$ are respectively the weights for $P,Q,R$. 

We have calculated this set of weights resorting to the point $S$, but it 
is clear that it can be obtained independently from the barycentric 
combinations
\begin{eqnarray}\label{weights}\hspace{-8mm}
&\displaystyle
P=\frac{\Omega_{Q}Q-\Omega_{W}W+\Omega_{R}R}{\Omega_{Q}-\Omega_{W}+
\Omega_{R}},\ 
Q=\frac{\Omega_{P}P-\Omega_{V}V+\Omega_{R}R}{\Omega_{P}-\Omega_{V}+
\Omega_{R}},&\nonumber\\ \hspace{-8mm} &\displaystyle
R=\frac{\Omega_{P}P-\Omega_{U}U+\Omega_{Q}Q}{\Omega_{P}-\Omega_{U}+
\Omega_{Q}},
&
\end{eqnarray}
up to a multiplicative factor. 

This is useful for degenerate quadrics (cones and cylinders), since
for their triangular quadric patches the boundary conics do not meet
in general at a point $S$, but we can still use (\ref{tconic}) as the
bilinear form for the tangent cone to the quadric along the conic on
$t$:

\begin{theorem}\label{degbilin}A rational triangular quadratic patch with control
net $\{c_{002}, 
c_{011},c_{020},c_{101},c_{110},c_{200}\}$ and weights 
$\{\omega_{002}, \omega_{011},\omega_{020},\omega_{101},\omega_{110},\omega_{200}
\}$, on a degenerate quadric has a bilinear form with coefficients $\Omega_{U}, \Omega_{V}, \Omega_{W}$ 
satisfying (\ref{weights}),
\begin{equation}\label{bilform1}\mathcal{Q}=
\frac{p^2}{\Omega_{W}^2}+\frac{q^2}{\Omega_{V}^2}+\frac{r^2}{\Omega_{U}^2}
-\frac{2pq}{\Omega_{W}\Omega_{V}}-\frac{2pr}{\Omega_{W}\Omega_{U}}
-\frac{2qr}{\Omega_{V}\Omega_{U}},\ 
\end{equation} 
with coefficients given by (\ref{Omegas}) and 
where $p$ is the linear form of the  plane containing
$c_{002}$, $c_{011}$, $c_{101}$, $q$ is the linear form of the plane
containing $c_{020}$, $c_{011}$, $c_{110}$ and $r$ is the linear form
of the plane containing $c_{200}$, $c_{101}$, $c_{110}$ which
satisfy
\[p(W)=1, \quad q(V)=1, \quad r(U)=1,\]
at the points $U$ (intersection of the planes $p,q,t$), $V$
 (intersection of the planes $p,r,t$) and $W$ (intersection of the 
planes $q,r,t$) given by (\ref{UVW}).

If the planes $p$, $q$, $r$ meet at a proper point $T$, the quadric 
is a cone with vertex $T$. If $T$ is a point at infinity, the quadric 
is a cylinder and the direction of its axis is given by $T$.
\end{theorem}

\section{Classification of quadrics}

The classification of quadric patches is refined now that we know whether
the quadric has a center or not:
\begin{itemize}
    \item  $\lambda>0$: Oval quadrics:
    \begin{itemize}
        \item  Centered: 
	\begin{itemize}
	\item  Ellipsoids: $\det \mathcal Z>0$.
	
	\item  Two-sheeted hyperboloids: $\det \mathcal Z<0$.
	\end{itemize}
    
        \item  Non-centered: Elliptic paraboloids.
    \end{itemize}
    
    \item $\lambda=0$: Degenerate quadrics:
    \begin{itemize}
        \item  Cones: if the vertex $T$ is a proper point.
    
        \item  Cylinders: if $T$ is a point at infinity. The type of 
	cylinder is determined classifying any of its conic sections 
	\cite{conics}. 
    \end{itemize}    

    \item  $\lambda<0$: Ruled quadrics:
    \begin{itemize}
        \item  Centered: One-sheeted hyperboloids.
    
        \item  Non-centered: Hyperbolic paraboloids.
    
    \end{itemize}

\end{itemize}

We may tell ellipsoids from two-sheeted hyperboloids in other ways. 
For instance, if the conics at planes $t,u,v,w$ are all ellipses, the 
quadric is an ellipsoid. We can use (\ref{canonical}) for this.

\section{Diametral planes and axes}

\begin{figure}
\centering
\includegraphics[height=4cm]{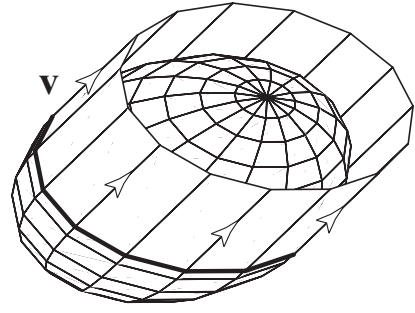}
\caption{The pole of a diametral plane is the direction of the 
tangent cylinder to the quadric along its intersection. }
\label{diametral}
\end{figure}

If a plane contains the center of the quadric, it is called 
\emph{diametral}. As the center of the quadric is the pole of the plane at 
infinity,  polar planes of points $\vec{v}$ at infinity are diametral. That 
is, the tangent cone to the quadric along its intersection with a 
diametral plane degenerates to a cylinder (see Fig.~\ref{diametral}). 
The direction of the cylinder is given by the pole $\vec{v}$ of the 
diametral plane.

We choose a basis of vectors $\{\vec U, \vec V, \vec W\}$, where 
\begin{equation}\label{rebase}\vec 
U=\overrightarrow{TU},\quad \vec V=\overrightarrow{TV},\quad \vec 
W=\overrightarrow{TW},\end{equation} if $U,V,W$ are proper points. If one of them 
is a point at infinity, we take it as vector of the basis. For 
instance, if $U$ is a point at infinity, we take $\vec{U}=U$. 

The case of an improper point $T$ can be handled similarly.  We use in
this case a basis $\{\vec U, \vec V, \vec T\}$ with $\vec
U=\overrightarrow{WU}$, $\vec V=\overrightarrow{WV}$, $\vec T=T$, if 
$W$ is a proper point. Otherwise, we choose $U$ or $V$ as origin and 
use $W$ as one of the vectors of the basis.

For a direction 
\begin{equation}\label{direction}\vec{v}=\alpha \vec{U}+ \beta\vec{V}+ 
\gamma \vec{W},\end{equation}
the polar plane is a diametral plane with linear form given by
\begin{eqnarray}\label{pdiametral}\hspace{-8mm}\mathcal 
Q\vec{v}\!\!\!\!\!&=&\!\!\!\!\!
\left(\frac{\gamma}{\Omega_{W}}-\frac{\alpha}{\Omega_{U}}-\frac{\beta}{\Omega_{V}}\right)
\frac{p}{\Omega_{W}}+\left(\frac{\beta}{\Omega_{V}}-\frac{\alpha}{\Omega_{U}}-\frac{\gamma}{\Omega_{W}}\right)
\frac{q}{\Omega_{V}}\nonumber\\\hspace{-8mm}\!\!\!\!\! &+&\!\!\!\!\!\!
\left(\frac{\alpha}{\Omega_{U}}-\frac{\beta}{\Omega_{V}}-\frac{\gamma}{\Omega_{W}}\right)
\!\frac{r}{\Omega_{U}}-\lambda 
(\epsilon_{U}\alpha+\epsilon_{V}\beta+\epsilon_{W}\gamma) t.\quad
\end{eqnarray}

Before going on, we need information about the normal of a plane 
given by a linear form:

\begin{lemma}A plane $l$ with linear form $ap+bq+cr+d t$, where $p,q,r,t$ 
 are linear forms satisfying
\[p(W)=q(V)=r(U)=t(T)=1,\] contains the vectors
\[\vec{v}_{1}=(b-\epsilon_{V}d)\vec{U}+(\epsilon_{U}d-c)\vec{V},\]\[ 
\vec{v}_{2}=(a-\epsilon_{W}d)\vec{V}+(\epsilon_{V}d-b)\vec{W},\]\[
\vec{v}_{3}=(\epsilon_{W}d-a)\vec{U}+(c-\epsilon_{U}d)\vec{W},\]
and hence its normal vector is given by
\begin{eqnarray*}\hspace{-7mm}
\vec{n}=(a-\epsilon_{W}d)\vec{U}\times\vec{V}+
(c-\epsilon_{U}d)\vec{V}\times\vec{W}+
(b-\epsilon_{V}d)\vec{W}\times\vec{U}.
\end{eqnarray*}
\end{lemma}

The proof is simple, since a vector $\vec{v}=\alpha \vec{U}+ \beta\vec{V}+ 
\gamma \vec{W}$ belongs to the plane if and only if
\[0=l(\vec{v})=(c-\epsilon_{U}d)\alpha+ (b-\epsilon_{V}d)\beta+ (a-
\epsilon_{W}d)\gamma.\]


A diametral plane is called a \emph{principal plane} or a \emph{plane
of symmetry} if it is orthogonal to its pole $\vec{v}$. The 
\emph{principal axis} of the quadric are the  lines which are 
intersections of two principal planes. That is, the poles of the 
principal planes are the directions of the axes.

Since for this definition we need to include a scalar product, it is 
necessary to provide another symmetric bilinear form $G$, such that 
$\vec{v}\cdot\vec{w}=G(\vec{v},\vec{w})$, denoting by a dot the 
scalar product.

The matrix of such form is 
usually called  Gram matrix and it is 
\begin{eqnarray}G=\left(
\begin{array}{ccc} g_{UU} &g_{UV} &g_{UW} \\
 g_{VU} &g_{VV} &g_{VW} \\ g_{WU} &g_{WV} &g_{WW}
\end{array}\right) \textrm{ where } g_{AB}:=\vec A\cdot \vec B
\label{gramian}
,\end{eqnarray}
in the basis 
$\{\vec{U},\vec{V},\vec{W}\}$.

In order to derive conditions for principal planes, we have to impose 
that the pole $\vec{v}$ be orthogonal to a basis of vectors of the 
diametral plane, which can be two of the ones which have been 
calculated in Lemma~1, 
\begin{eqnarray*}
0&=&\vec{v}_{1}\cdot  \vec{v}=(b-\epsilon_{V}d)(g_{UU}\alpha+g_{UV}\beta+g_{UW}\gamma)\\&+&
(\epsilon_{U}d-c)(g_{VU}\alpha+g_{VV}\beta+g_{VW}\gamma),\\
0&=&\vec{v}_{2}\cdot  \vec{v}=(a-\epsilon_{W}d)(g_{VU}\alpha+g_{VV}\beta+g_{VW}\gamma)\\&+& 
(\epsilon_{V}d-b)(g_{WU}\alpha+g_{WV}\beta+g_{WW}\gamma),
\end{eqnarray*}
and these equations can be easily solved up to a proportionality 
factor $\mu$,
\begin{eqnarray}\label{musystem}
a-\epsilon_{W}d&=&\mu (g_{WU}\alpha+g_{WV}\beta+g_{WW}\gamma),\nonumber\\
b-\epsilon_{V}d&=&\mu(g_{VU}\alpha+g_{VV}\beta+g_{VW}\gamma),\nonumber\\
c-\epsilon_{U}d&=&\mu(g_{UU}\alpha+g_{UV}\beta+g_{UW}\gamma),
\end{eqnarray}
where
\begin{eqnarray}\label{ortho}
a&=&\left(\frac{\gamma}{\Omega_{W}}-\frac{\alpha}{\Omega_{U}}-\frac{\beta}{\Omega_{V}}\right)
\frac{1}{\Omega_{W}},\nonumber\\
b&=&\left(\frac{\beta}{\Omega_{V}}-\frac{\alpha}{\Omega_{U}}-\frac{\gamma}{\Omega_{W}}\right)
\frac{1}{\Omega_{V}},\nonumber\\
c&=&\left(\frac{\alpha}{\Omega_{U}}-\frac{\beta}{\Omega_{V}}-\frac{\gamma}{\Omega_{W}}\right)
\frac{1}{\Omega_{U}},\nonumber\\
d&=&-\lambda (\epsilon_{U}\alpha+\epsilon_{V}\beta+\epsilon_{W}\gamma).
\end{eqnarray}

In the case of improper $T$ and, for instance, proper $W$, the equations for the coordinates of
$\vec{v}=\alpha \vec U+\beta \vec V+\delta \vec T$ are
\begin{eqnarray}\label{tsystem}
b-\epsilon_{V}a&=&\mu (g_{VU}\alpha+g_{VV}\beta+g_{VT}\delta),\nonumber\\
c-\epsilon_{U}a&=&\mu(g_{UU}\alpha+g_{UV}\beta+g_{UT}\delta),\nonumber\\
d&=&\mu(g_{TU}\alpha+g_{TV}\beta+g_{TT}\delta),
\end{eqnarray}
where
\begin{eqnarray}\label{tortho}
a&=&\left(-\frac{\epsilon_{U}\alpha+\epsilon_{V}\beta}{\Omega_{W}}-\frac{\alpha}{\Omega_{U}}-\frac{\beta}{\Omega_{V}}\right)
\frac{1}{\Omega_{W}},\nonumber\\
b&=&\left(\frac{\beta}{\Omega_{V}}-\frac{\alpha}{\Omega_{U}}+\frac{\epsilon_{U}\alpha+\epsilon_{V}\beta}{\Omega_{W}}\right)
\frac{1}{\Omega_{V}},\nonumber\\
c&=&\left(\frac{\alpha}{\Omega_{U}}-\frac{\beta}{\Omega_{V}}+\frac{\epsilon_{U}\alpha+\epsilon_{V}\beta}{\Omega_{W}}\right)
\frac{1}{\Omega_{U}},\nonumber\\
d&=&\lambda \delta.
\end{eqnarray}

These conditions can be seen as arising from an alternative definition
of principal axes as lines with direction given by eigenvectors of the
bilinear form of the conic at infinity.  The values of the coefficient
$\mu$ are the corresponding eigenvalues, which are obtained by
imposing that the system (\ref{musystem}) has non-trivial solutions
for $\alpha,\beta,\gamma$.  Hence $\mu$ has to satisfy a cubic
equation and there are in general three principal planes and axes,
except for quadrics of revolution and spheres:

\begin{corollary}\label{planes}
A rational triangular quadratic patch with control net $\{c_{002},
c_{011},c_{020},c_{101},c_{110},c_{200}\}$ and weights
$\{\omega_{002},
\omega_{011},\omega_{020},\omega_{101},\omega_{110},\omega_{200} \}$,
such that the three boundary conics meet at a point $S$, which is
written as in (\ref{baryS}), has diametral planes with linear forms
given by (\ref{pdiametral}) with a pole $\vec{v}=\alpha
\vec{U}+\beta\vec{V}+\gamma
\vec{W}$, or $\vec{v}=\alpha
\vec{U}+\beta\vec{V}+\gamma
\vec{T}$, if $T$ is a point at infinity, in a vector basis 
(\ref{rebase}), defined with points 
(\ref{UVW}) and (\ref{TTT}).

The poles of principal planes are the directions of principal axes 
and, if $T$ is a proper 
point, they satisfy the linear system (\ref{musystem}) for values of $\mu$ for which the determinant
\[\hspace{-8mm}
\scriptsize{\left|\begin{array}{ccc}\hspace{-2mm}
\lambda_{WU}-\frac{1}{\Omega_{W}\Omega_{U}}-\mu g_{WU}  & 
\hspace{-3mm}
\lambda_{WV}-\frac{1}{\Omega_{W}\Omega_{V}}-\mu g_{WV} &
\hspace{-3mm}
\lambda_{W}+\frac{1}{\Omega_{W}^2}-\mu g_{WW} \hspace{-3mm} \\[2mm] \hspace{-3mm}
\lambda_{VU}-\frac{1}{\Omega_{V}\Omega_{U}}-\mu g_{VU}  & 
\hspace{-3mm}
\lambda_{V}+\frac{1}{\Omega_{V}^2}-\mu g_{VV} &
\hspace{-3mm}
\lambda_{VW}-\frac{1}{\Omega_{V}\Omega_{W}}-\mu g_{VW}\hspace{-2mm}  \\[2mm]\hspace{-3mm}
\lambda_{U}+\frac{1}{\Omega_{U}^2}-\mu g_{UU}  &
\hspace{-3mm}
\lambda_{UV}-\frac{1}{\Omega_{U}\Omega_{V}}-\mu g_{UV} &
\hspace{-3mm}
\lambda_{UW}-\frac{1}{\Omega_{U}\Omega_{W}}-\mu g_{UW}   \hspace{-2mm}  
\end{array}
\right|}\]
vanishes, with $\lambda_{AB}:=\epsilon_{A}\epsilon_{B}\lambda$ and 
coefficients given by (\ref{Omegas}) and (\ref{gramian}).

If $T$ is a point at infinity, the coordinates of the pole satisfy 
the linear system (\ref{tsystem}) for values of $\mu$ for which the 
determinant
\[
\left|\begin{array}{ccc}
A_{UU}-\mu g_{UU} & 
A_{UV}-\mu g_{UV} &
\hspace{-2mm}-\mu g_{UT} \\
A_{UV}-\mu g_{VU} &
A_{VV}-\mu g_{VV} & 
\hspace{-2mm}-\mu g_{VT} \\
-\mu g_{TU} & -\mu g_{TV} & \lambda-\mu g_{TT}
\end{array}\right|
\]
vanishes, with
\[A_{UU}:=\left(\frac{\epsilon_{U}}{\Omega_{W}}+\frac{1}{\Omega_{U}}\right)^2,\quad
A_{VV}:=\left(\frac{\epsilon_{V}}{\Omega_{W}}+\frac{1}{\Omega_{V}}\right)^2,\]
\[A_{UV}:=\frac{\epsilon_{U}\epsilon_{V}}{\Omega_{W}^2}+\frac{\epsilon_{V}}{\Omega_{W}\Omega_{U}}+
\frac{\epsilon_{U}}{\Omega_{W}\Omega_{V}}-\frac{1}{\Omega_{U}\Omega_{V}}.\]

In general there are three different values for $\mu$. If there is a 
double non-null solution, the quadric is a surface of revolution. If there is 
a triple solution, the surface is a sphere.
\end{corollary}

The discriminant of a cubic equation 
$a_{3}\mu^3+a_{2}\mu^2+a_{1}\mu+a_{0}=0$,
\[\Delta=-27a_{3}^2a_{0}^2+18a_{0}a_{1}a_{2}a_{3}-4a_{3}a_{1}^3-4a_{2}^3a_{0}+a_{2}^2a_{1}^2,\]
provides a simple way of checking whether a quadric is a surface of 
revolution: A vanishing discriminant $\Delta$ is equivalent to 
having a double root.

Besides, a vanishing second derivative of the equation, 
$3a_{3}\mu+2a_{2}=0$, 
implies a triple root and hence the quadric would be a sphere.

It is easily checked that the eigenvalue $\mu=0$ just appears for 
$\lambda^{-1}=\tilde\Omega_{U}\tilde\Omega_{V}+\tilde\Omega_{V}\tilde\Omega_{W}+\tilde\Omega_{W}\tilde\Omega_{U}$ in the case of proper $T$. 
That is, for paraboloids. For these surfaces the plane at infinity 
is a principal plane: The plane at infinity is diametral, as it 
comprises the center, and it is principal, since it fulfills 
(\ref{musystem}). Besides, it is the polar plane of the center, since it is 
tangent to the paraboloid at the center.
\begin{figure}
\centering
\includegraphics[height=4cm]{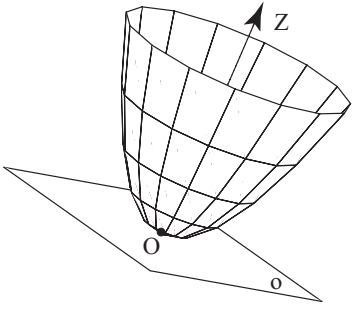}
\caption{The vertex $O$ of a paraboloid is the pole of a tangent 
plane with the center $Z$ as normal. }
\label{vertice}
\end{figure}

In the case of improper $T$, the eigenvalue $\mu=0$ appears only if
$\lambda=0$.  Hence, cylinders have a null eigenvalue, corresponding
to a pole $T$, the direction of the axis, which has no polar plane.
For parabolic cylinders there is another null eigenvalue, since the 
plane at infinity is an improper principal plane.

There are then just two proper principal planes 
for paraboloids, except for paraboloids of revolution. The 
intersection of these principal planes is the only proper \emph{axis} of the 
paraboloid, a line with direction given by the center of the 
paraboloid. The axis meets the paraboloid at the center and at a single proper point 
named \emph{vertex}. 


We may calculate the vertex solving a quadratic equation, but there is
a simpler way, taking into account that the tangent plane $o$ at the
vertex $O$ is orthogonal to the center $Z$ (see Fig.~\ref{vertice}).
We can use this property to compute the vertex:

A plane with linear form $ap+bq+cr+dt$ is orthogonal to a vector 
(\ref{direction}) if their coordinates fulfill (\ref{musystem}). 
Hence reading the coordinates of  $Z$ from (\ref{centerpole}) with, 
for instance, $\mu=1$,
\[\alpha=\Omega_{U}(\tilde\Omega_{V}+\tilde\Omega_{W}),\quad 
\beta=\Omega_{V}(\tilde\Omega_{W}+\tilde\Omega_{U}),\]\[ 
\gamma=\Omega_{W}(\tilde\Omega_{U}+\tilde\Omega_{V}),\]we get the 
differences $A=a-\epsilon_{W}d$, $B=b-\epsilon_{V}d$, $C=c-\epsilon_{U}d$ for the planes which are orthogonal 
to the center. That is, we have the linear form for the tangent plane at the vertex of 
the paraboloid, except for the coefficient $d$. Since tangent planes 
are solutions of the implicit equation of the quadric in tangential 
form (\ref{tangential}),
\begin{eqnarray*}\hspace{-7mm}
0\!\!\!&=&\!\!\!\Omega_{U}\Omega_{V}(C+\epsilon_{U}d)(B+\epsilon_{V}d)
\\\hspace{-7mm}
\!\!\!&+&\!\!\!
\Omega_{V}\Omega_{W}(B+\epsilon_{V}d)(A+\epsilon_{W}d)\\\hspace{-7mm}
\!\!\!&+&\!\!\!
\Omega_{W}\Omega_{U}(A+\epsilon_{W}d)(C+\epsilon_{U}d)
\\\hspace{-7mm}
\!\!\!&-&\!\!\!
\frac{1}{\tilde\Omega_{U}\tilde\Omega_{V}+
\tilde\Omega_{V}\tilde\Omega_{W}+\tilde\Omega_{W}\tilde\Omega_{U}}\\\hspace{-7mm}
\!\!\!&=&\!\!\!
\Omega_{U}\Omega_{V}BC+\Omega_{V}\Omega_{W}AB+
\Omega_{W}\Omega_{U}AD\\\hspace{-7mm}
\!\!\!&+&\!\!\!
\big(\Omega_{U}\Omega_{V}(\epsilon_{U}B+\epsilon_{V}C)+
\Omega_{V}\Omega_{W}(\epsilon_{V}A+\epsilon_{W}B)
\\\hspace{-7mm}
\!\!\!&+&\!\!\!
\Omega_{W}\Omega_{U}(\epsilon_{U}A+\epsilon_{W}C)\big)d
,\end{eqnarray*}
the coefficients of the plane are readily obtained
\begin{eqnarray}\label{pole}
\hspace{-8mm}d\!\!\!\!&=&\!\!\!\!-
\frac{\Omega_{U}\Omega_{V}BC+\Omega_{V}\Omega_{W}AB+
\Omega_{W}\Omega_{U}AC}{\scriptstyle{\Omega_{U}\Omega_{V}(\epsilon_{U}B+\epsilon_{V}C)+
\Omega_{V}\Omega_{W}(\epsilon_{V}A+\epsilon_{W}B)+
\Omega_{W}\Omega_{U}(\epsilon_{U}A+\epsilon_{U}C)}},\nonumber\\ \hspace{-8mm}
a\!\!\!\!&=&\!\!\!\! \epsilon_{W}
d+g_{WU}\Omega_{U}(\tilde\Omega_{V}+\tilde\Omega_{W})+
g_{WV}\Omega_{V}(\tilde\Omega_{W}+\tilde\Omega_{U})\nonumber\\ 
\hspace{-8mm} \!\!\!\!&+&\!\!\!\!
g_{WW}\Omega_{W}(\tilde\Omega_{U}+\tilde\Omega_{V}),\nonumber\\ \hspace{-8mm}
b\!\!\!\!&=&\!\!\!\!\epsilon_{V}
d+g_{VU}\Omega_{U}(\tilde\Omega_{V}+\tilde\Omega_{W})+
g_{VV}\Omega_{V}(\tilde\Omega_{W}+\tilde\Omega_{U})\nonumber\\ 
\hspace{-8mm} \!\!\!\!&+&\!\!\!\!
g_{VW}\Omega_{W}(\tilde\Omega_{U}+\tilde\Omega_{V}),\nonumber\\ \hspace{-8mm}
c\!\!\!\!&=&\!\!\!\!\epsilon_{U}
d+g_{UU}\Omega_{U}(\tilde\Omega_{V}+\tilde\Omega_{W})+
g_{UV}\Omega_{V}(\tilde\Omega_{W}+\tilde\Omega_{U})\nonumber\\ 
\hspace{-8mm} \!\!\!\!&+&\!\!\!\!
g_{UW}\Omega_{W}(\tilde\Omega_{U}+\tilde\Omega_{V}),
\end{eqnarray}
where $a,b,c,d$ are given by (\ref{ortho}).

The vertex $O$ is just the pole of the plane $o=ap+bq+cr+dt$:
\begin{corollary}
A rational triangular quadratic patch for a paraboloid with control net $\{c_{002},
c_{011},c_{020},c_{101},c_{110},c_{200}\}$ and weights
$\{\omega_{002},
\omega_{011},\omega_{020},\omega_{101},\omega_{110},\omega_{200} \}$,
such that the three boundary conics meet at a point $S$, which is
written as in (\ref{baryS}), has a vertex given by
\begin{eqnarray*}\label{vertex}\hspace{-8mm}
O\!\!\!\!\!&=&\!\!\!\!\!\frac{(b\Omega_V+a\Omega_W)\Omega_UU+(c\Omega_U+a\Omega_W)\Omega_VV}
{\scriptstyle{\Omega_{U}\Omega_{V}(b+c-2\epsilon_{U}\epsilon_{V}d)+
\Omega_{V}\Omega_{W}(a+b-2\epsilon_{V}\epsilon_{W}d)+
\Omega_{W}\Omega_{U}(a+c-2\epsilon_{W}\epsilon_{U}d)}}
\\\hspace{-8mm}\!\!\!\!\!&+&\!\!\!\!\!
\frac{(c\Omega_U+b\Omega_V)\Omega_WW-2d(\tilde\Omega_{U}\tilde\Omega_{V}+
\tilde\Omega_{V}\tilde\Omega_{W}+\tilde\Omega_{W}\tilde\Omega_{U})T}
{\scriptstyle{\Omega_{U}\Omega_{V}(b+c-2\epsilon_{U}\epsilon_{V}d)+
\Omega_{V}\Omega_{W}(a+b-2\epsilon_{V}\epsilon_{W}d)+
\Omega_{W}\Omega_{U}(a+c-2\epsilon_{W}\epsilon_{U}d)}},
\end{eqnarray*}
where $A=a-\epsilon_{W}d$, $B=b-\epsilon_{V}d$, $C=c-\epsilon_{U}d$
and (\ref{pole}).  The coefficients are given in (\ref{Omegas}) and
(\ref{tildeOmegas}) and the points $U,V,W,T$ are defined in
(\ref{UVW}) and (\ref{TTT}).
\end{corollary}

Finally, the axis of a cylinder is easily determined, since it has 
the direction of $T$ and contains the center of every conic section. 
For instance, we can use the center of the conic at $t$, which 
according to (\ref{center}) is 
given by \[\hspace{-5mm}
Z_{t}=\frac{\epsilon_{W}\Omega_{W}\omega_{w}\omega_{002}P+
\epsilon_{V}\Omega_{V}\omega_{v}\omega_{020}Q+
\epsilon_{U}\Omega_{U}\omega_{u}\omega_{200}R}{\Omega_{Z}}
.\]
\section{Examples}

\begin{figure}
\centering
\includegraphics[height=6cm]{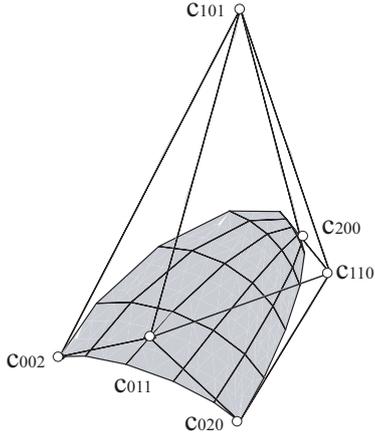}
\caption{Ellipsoid}
\label{elipsoide}
\end{figure}

Now we apply our results to several quadric patches:

\begin{example} Net: 
$\left[\begin{array}{c}(0,0,0)\hspace{0.2cm}(1,0,1)
\hspace{0.2cm}(2,0,0)\\(-2,2,3)\hspace{0.2cm}(1,1,1)\\ (0,\frac{4}{3},\frac{2}{3})\end{array}\right]$ and  
 weights: 
 $\left[\begin{array}{c}1\hspace{0.2cm}\frac{1}{2}\hspace{0.2cm}1\\\frac{1}{3}
 \hspace{0.1cm}\frac{2}{3}\\ 
1\end{array}\right]$ (Fig.~\ref{elipsoide}):\end{example}

The normalised linear forms for the tangent planes are 
\[\hspace{-5mm}
p=\frac{10x+25y-10z}{32},\ q=\frac{2-x-z}{8},\ r=
\frac{12-3x-9y}{16},\]and meet at a point $(-8/7, 12/7, 22/7)$.

The three boundary conics meet at a point,
\begin{eqnarray*}\hspace{-5mm}
S&=&(1, 0, -1)=c_{002}-c_{011}+c_{020}\\&=&
\frac{c_{002}-2c_{101}/3+c_{200}}{4/3}=\frac{c_{200}-4c_{110}/3+c_{020}}{2/3},\end{eqnarray*}
and hence $\omega_{u}=1$, $\omega_{v}=4/3$, $\omega_{w}=2/3$.

The normalised linear form for the plane through the corners of the 
net is 
\[ t=\frac{-7y+14z}{32}.\]

The quadric is oval, since the coefficient  $\lambda=207/49$ for this 
quadric patch is positive.

It is not a paraboloid, since the center is the proper point $(1/2,
2/5, -1/30)$.  Since the boundary curves and the conic on $t$ are
ellipses, the quadric is an ellipsoid.  One arrives to the same
conclusion checking that $\det\mathcal Z=480/49$ is positive.

The bilinear form that we get for this surface is
\[\hspace{-7mm}\mathcal 
Q=\frac{18x^2+45xy+81y^2-81yz+54z^2-36x-90y+36z}{64},\] and the implicit equation, in cartesian 
coordinates is
\[\hspace{-7mm}0=18x^2+45xy+81y^2-81yz+54z^2-36x-90y+36z.\]

The three eigenvalues calculated according to Corollary~\ref{planes} 
are different, $\mu=1.78, 0.90, 0.52$, and so this ellipsoid is not a 
surface of revolution.

The three principal planes are
\[ 0.61 p + 0.37 q - 1.32 r - 2.04 t, \]\[
 -0.90 p + 15.05 q - 4.49 r + 5.04 t,\]\[
 -10.46 p + 17.24 q - 0.57 r - 19.12 t,\]
with respective implicit equations in cartesian coordinates
\[ 0.39 x + 1.67 y - 1.13 z - 0.90=0,\]
\[  -1.32 x + 0.72 y + 0.60 z + 0.39=0,\]
\[  -5.32 x - 3.67 y - 7.25 z + 3.88=0,\]
and obviously meet at the center.

\begin{figure}
\centering
\includegraphics[height=6cm]{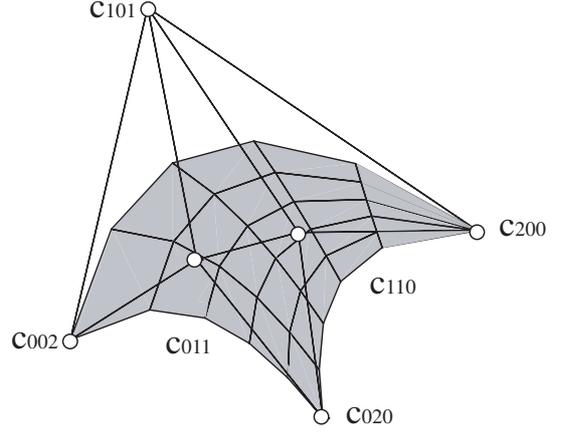}
\caption{Two-sheeted hyperboloid}
\label{hyperboloid}
\end{figure}

\begin{example} Net: 
$\left[\begin{array}{c}(0,0,0)\hspace{0.2cm}(1,0,1)
\hspace{0.2cm}(2,0,0)\\(-1,1,2)\hspace{0.2cm}(1,\frac{1}{2},1)\\ (0,2,0)\end{array}\right]$ and  
 weights: 
 $\left[\begin{array}{c}1\hspace{0.2cm}2\hspace{0.2cm}3\\1\hspace{0.2cm}2\\ 
1\end{array}\right]$ (Fig.~\ref{hyperboloid}):\end{example}

The normalised linear forms for the tangent planes are 
\[\hspace{-5mm}
p=\frac{x+3y-z}{4},\ q=\frac{x+z-2}{4},\ r=
\frac{12-4x-6y-5z}{8},\]and meet at a point $(2/5, 2/5, 8/5)$.

The three boundary conics meet at a point at infinity,
\begin{eqnarray*}\hspace{-5mm}
S&=&(2, 0, -4)=c_{002}-4c_{011}+3c_{020}\\&=&
c_{002}-2c_{101}+c_{200}=c_{200}-4c_{110}+3c_{020},\end{eqnarray*}
and hence $\omega_{u}=\omega_{v}=\omega_{w}=1$.

The normalised linear form for the plane through the corners of the 
net is 
\[ t=\frac{5z}{8}.\]

The quadric is oval, since the coefficient  $\lambda=3/25$ for this 
quadric patch is positive.

It is not a paraboloid, since the center is the proper point
$(7/4,-1/2,5/2)$.  As the boundary curves are not ellipses, but two
hyperbolas and one parabola, the quadric is not an ellipsoid, but a
two-sheeted hyperboloid. Accordingly,  $\det\mathcal Z=-16/25$ is negative.

The bilinear form that we get for this surface is
\[\mathcal Q=\frac{4x^2+12xy+12y^2+6yz-z^2-8x-24y+8z}{48},\] and the implicit equation, in cartesian 
coordinates is
\[0=4x^2+12xy+12y^2+6yz-z^2-8x-24y+8z.\]

The eigenvalues for the normal directions of the principal planes 
are different, $\mu=0.33, -0.44, 0.30$, and hence the hyperboloid is 
not a surface of revolution.

The principal planes are
\[ 1.58 p + 0.45 q - 0.91 r  + 0.08 t,\]
\[0.36 p + 0.05 q + 0.30 r + 0.22 t,\]
\[0.79 p + 4.65 q + 1.21 r - 0.96 t,\]
and have the respective equations in cartesian coordinates
\[ 0.97 x + 1.87 y + 0.34 z  - 1.60=0,\]
\[-0.05 x + 0.05 y - 0.13 z + 0.43=0,\]
\[ 0.75 x - 0.32 y - 0.39 z - 0.51=0,\]
and meet at the center as expected.

\begin{figure}
\centering
\includegraphics[height=3.5cm]{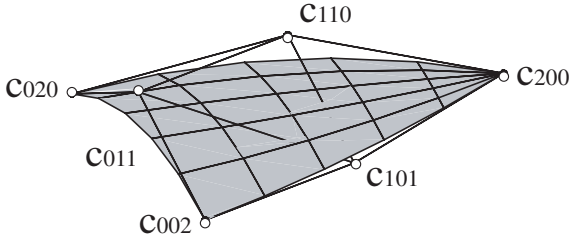}
\caption{Hyperbolic paraboloid}
\label{hyperparaboloid}
\end{figure}

\begin{example} Net: 
$\left[\begin{array}{c}(0,0,0)\hspace{0.2cm}(1,0,0)\hspace{0.2cm}(2,0,2)\\(0,\frac{1}{2},0)\hspace{0.2cm}(1,\frac{1}{2},0)\\
(0,1,-\frac{1}{2})\end{array}\right]$ and weights: $\left[\begin{array}{c}1 
\hspace{0.2cm} 1 \hspace{0.2cm} 1\\
1 \hspace{0.2cm} 1\\ 1\end{array}\right]$ (Fig.~\ref{hyperparaboloid})\end{example}

The normalised linear forms for the tangent planes are 
\[
p=-z,\ q=\frac{4-4x+2z}{3},\ r=
\frac{-1+2y+2z}{3},\]and meet at a point $(1, 1/2, 0)$.

The three boundary conics meet at a point at infinity,
\begin{eqnarray*}\hspace{-5mm}
S&=&(0, 0, 2)=c_{002}-2c_{011}+c_{020}\\&=&
\frac{c_{002}-2c_{101}+c_{200}}{-1/4}=\frac{c_{200}-2c_{110}+c_{020}}{3/4},\end{eqnarray*}
and hence $\omega_{u}=1$, $\omega_{v}=-1/4$, $\omega_{w}=3/4$.

The normalised linear form for the plane through the corners of the 
net is 
\[ t=\frac{4x-2y-4z}{3}.\]

The quadric is ruled, since the coefficient  $\lambda=-1$ for this 
quadric patch is negative.

It is a hyperbolic paraboloid, since the center is a point at infinity
$(0,0,-3/64)$, which is also the direction of the axis.

The bilinear form that we get for this surface is
\[\mathcal Q=\frac{8z-4x^2+4y^2}{3},\] and the implicit equation, in cartesian 
coordinates is
\[0=2z-x^2+y^2.\]

The eigenvalues calculated according to Corollary~\ref{planes} 
are different, $\mu=0, 4/3, -4/3$, and one of them is null, as it is 
expected for a hyperbolic paraboloid.

The three principal planes are
\[q + r + t, \]\[
 2p+q+4r+t,\]\[
 -\frac{3}{2}p+\frac{3}{4}q+3r+3t.\]
 
The first principal plane is the plane at infinity and the other
ones have $y=0$, $x=0$ as implicit equations in cartesian 
coordinates. They all meet at the center. 

The vertex is the point $(0,0,0)$, as it is clear from the form of 
the implicit equation.

\begin{figure}
\centering
\includegraphics[height=5cm]{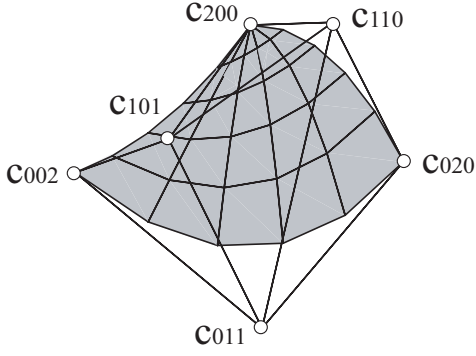}
\caption{Parabolic cylinder}
\label{paracilindro}
\end{figure}

\begin{example} Net: 
$\left[\begin{array}{c}(0,-1,0)\hspace{0.1cm}(2,0,0)
\hspace{0.1cm}(0,1,0)\\(1,-\frac{1}{2},\frac{1}{2})\hspace{0.1cm}(1,\frac{1}{2},1)\\ (1,0,1)\end{array}\right]$ and  
 weights: 
 $\left[\begin{array}{c}1\hspace{0.2cm}1\hspace{0.2cm}1\\1\hspace{0.2cm}1\\ 
1\end{array}\right]$ (Fig.~\ref{paracilindro}):\end{example}

The normalised linear forms for the tangent planes are 
\[\hspace{-5mm}
p=1-\frac{x}{2}+y,\ q=1-\frac{x}{2}-y,\ r=
x-1,\]and meet at a point at infinity $(0,0,1)$.

The three boundary conics are parabolas and do not meet at any point. 
Hence the patch does not belong to a non-degenerate quadric.  If
it is a degenerate quadric, it is then a parabolic cylinder with direction 
$(0,0,1)$. It is easy to check, for instance, that their respective 
\emph{centers} are aligned and hence it is a degenerate quadric.

The normalised linear form for the plane through the corners of the 
net is 
\[ t=-x+z.\]

Using (\ref{weights}) we find a set of weights $\Omega_{U}=-1/2$, 
$\Omega_{V}=1=\Omega_{W}$ and hence the bilinear form that we get for this surface is
\[\mathcal Q=4(x+y^2-1),\] and the implicit equation, in cartesian 
coordinates is
\[x=1-y^2.\]

The eigenvalues for the poles of the principal planes 
are $\mu=0,0,4$, as it is expected for a parabolic cylinder.

The principal planes are
\[p+q+r,\qquad p - q.\]

The first one is the plane at infinity and the second one is the only
proper principal plane of a parabolic cylinder, with equation in
cartesian coordinates given by $y=0$.

\begin{figure}
\centering
\includegraphics[height=5cm]{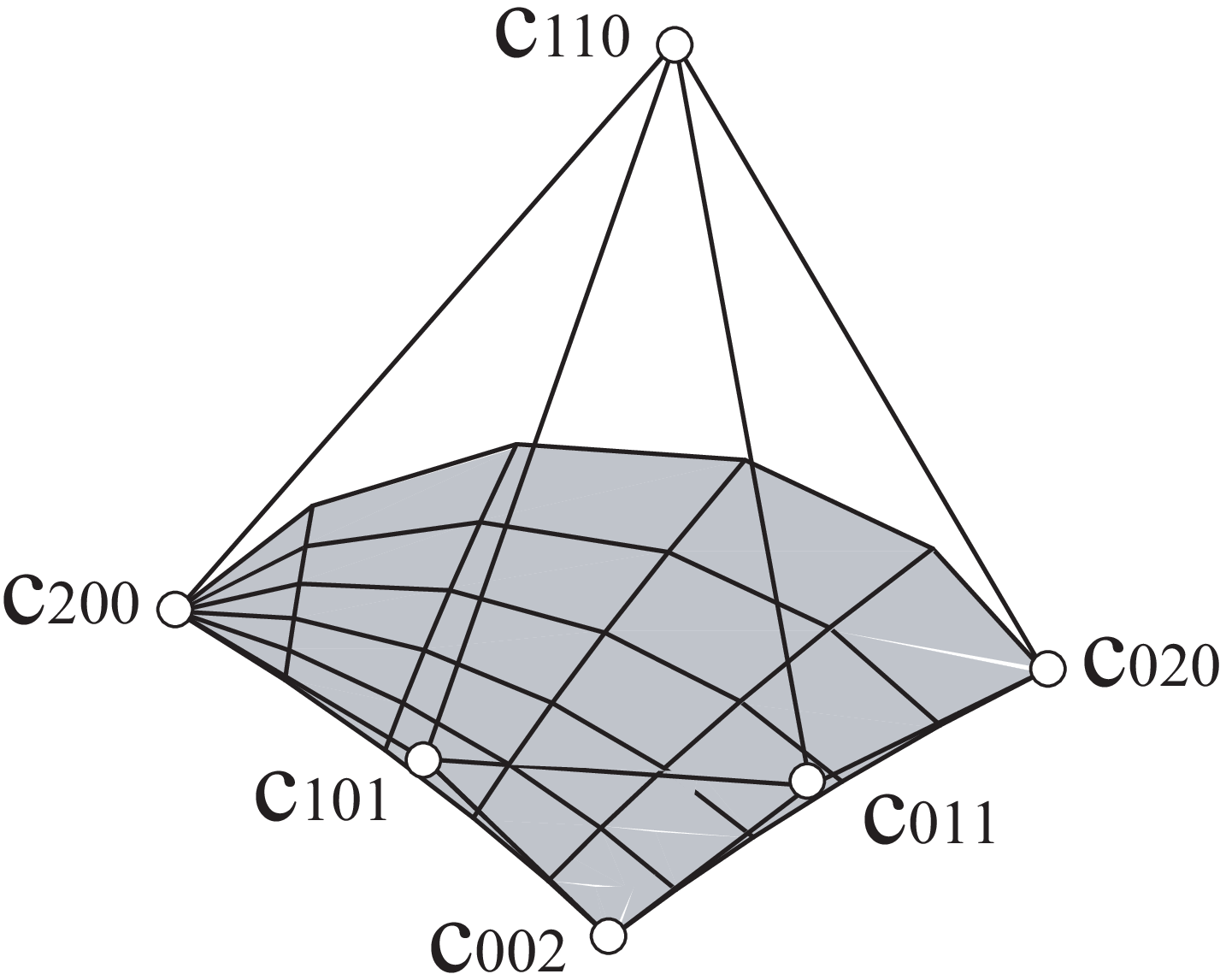}
\caption{Cone}
\label{cono}
\end{figure}

\begin{example} Net:%
$\left[\begin{array}{c}(2,0,2)\hspace{0.5mm}(\frac{4}{3},\frac{4\sqrt{3}}{9},\frac{4}{3})
\hspace{0.5mm}(\frac{1}{2},\frac{\sqrt{3}}{2},1)
\\[1mm](\frac{4}{3},-\frac{4\sqrt{3}}{9},\frac{4}{3})\hspace{0.5mm}(2,0,1)
\\[1mm] 
(\frac{1}{2},-\frac{\sqrt{3}}{2},1)\end{array}\right]$ and  
 weights: 
$\left[\begin{array}{c}\frac{3}{8}\hspace{0.2cm}\frac{9}{16}\hspace{0.2cm}1\\
\frac{9}{16}\hspace{0.2cm}\frac{1}{2}\\ 
1\end{array}\right]$ (Fig.~\ref{cono}):\end{example}

The normalised linear forms for the tangent planes are 
\[\hspace{-5mm}
p=\frac{z-x}{2},\ q=\frac{-x-\sqrt{3}y+2z}{4},\ r=\frac{-x+\sqrt{3}y+2z}{4}
,\]and meet at a point $(0,0,0)$.

The three boundary conics meet at a point
\begin{eqnarray*}\hspace{-5mm}
S&=&(-1, 0, 1)=\frac{3c_{002}/8-9c_{011}/8+c_{020}}{1/4}\\&=&
\frac{3c_{002}/8-9c_{101}/8+c_{200}}{1/4}=c_{200}-c_{110}+c_{020},\end{eqnarray*}
and hence $\omega_{u}=1/4=\omega_{v}$, $\omega_{w}=1$.

The normalised linear form for the plane through the corners of the 
net is 
\[ t=\frac{2+2x-3z}{2}.\]

The quadric is degenerate, since the coefficient for this quadric
patch is $\lambda=0$.  It is a cone, since the intersection
of the tangent planes is a proper point, which is the vertex $(0,0,0)$.

The bilinear form that we get for this surface is
\[\mathcal Q=\frac{16}{3}(x^2+y^2-z^2),\] and the implicit equation, in cartesian 
coordinates is
\[0=x^2+y^2-z^2.\]

The eigenvalues for the poles of the principal planes 
are $\mu=-16/3, 16/3, 16/3$, and so the surface is a cone of 
revolution.

The principal planes are
\[  p - q - r,\]
\[3\gamma p+(1-2\gamma) q-(1+\gamma)r,\]
for every value of $\gamma$ 
and  their respective equations in cartesian coordinates are
\[ z=0,\]
\[3\gamma x+(2-\gamma)\sqrt{3}y=0.\]

\section{Tensor product quadric patches}

Tensor product patches are the most common way to model surfaces in 
CAD. In particular, in some cases quadrics can be parametrised by 
biquadratic rational B\'ezier patches,
\[c(u,v)=\frac{\displaystyle
\sum_{i=0}^{2}\sum_{j=0}^2\omega_{ij}c_{ij}B^{2}_{i}(u)B^2_{j}(v)}
{\displaystyle\sum_{i=0}^{2}\sum_{j=0}^2\omega_{ij}B^{2}_{i}(u)B^2_{j}(v)},\quad 
u,v\in[0,1],\]
for a control net 
$\{c_{00},c_{01},c_{02},c_{10},c_{11},c_{12},c_{20},c_{21},c_{22}\}$ 
and their respective weights.

The patch is bounded by four conic arcs with control polygons 
$\{c_{00},c_{01},c_{02}\}$, $\{c_{20},c_{21},c_{22}\}$, 
$\{c_{00},c_{10},c_{20}\}$ and $\{c_{02},c_{12},c_{22}\}$, meeting 
two by two at the four corner vertices $c_{00}$, $c_{02}$, $c_{20}$, 
$c_{22}$.

Not every rational biquadratic patch is a quadric patch 
\cite{boehm-quadrics, fink}, but we can apply our knowledge about quadric 
triangular patches to them. 

\begin{figure}
\centering
\includegraphics[height=5cm]{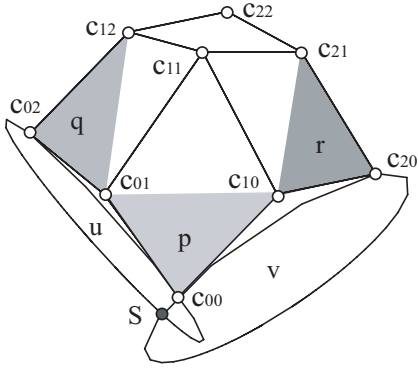}
\caption{Planes on a biquadratic tensor product patch}
\label{tensorial}
\end{figure}

For instance, we can take $P=c_{00}$, 
$Q=c_{02}$, $R=c_{20}$ and define a triangular patch with these three 
corners, as we know that the tangent planes are defined by the 
neighbouring vertices: $p$ contains $c_{00},c_{01},c_{10}$, $q$ 
contains $c_{02},c_{01},c_{12}$ and $r$ contains $c_{20}$, $c_{10}$, 
$c_{21}$ (see Fig.~\ref{tensorial}).

We already know the conic at $u$, defined by the control points 
$\{c_{00},c_{01},c_{02}\}$ and their weights, and the conic at $v$, defined by  
$\{c_{00},c_{10},c_{20}\}$ and their weights. For the conic at $w$ we 
have the control points $c_{02}$ and $c_{20}$ and their weights, but 
we lack the intermediate control point $c_{110}$ and the weight 
$\omega_{110}$.

In order to have a triangular quadric patch, we can use the other point $S$
where the conics at $u$ and $v$ meet, besides $P$.  We may
reparametrise both conics as we did in (\ref{baryS}) so that their
weights satisfy \begin{eqnarray*}\hspace{-8mm}S=
\frac{\omega_{00}c_{00}-2\omega_{01}c_{01}+
\omega_{02}c_{02}}{\omega_{u}} =
\frac{\omega_{00}c_{00}-2\omega_{10}c_{10}+\omega_{20}c_{20}}{\omega_{v}},
\end{eqnarray*}
where denominators, if $S$ is a proper point, are
\begin{eqnarray*}
\omega_{u}=\omega_{00}-2\omega_{01}+ \omega_{02},\quad 
\omega_{v}=\omega_{00}-2\omega_{10}+ \omega_{20}.
\end{eqnarray*}

Now we can define the plane $w$ as the one containing 
$S,c_{02},c_{20}$ and complete Fig.~\ref{circumscribe} by computing 
the intersection points $U,V,W$ on plane $t$.

The barycentric combinations for $U,V,W$ provide us the value of 
$\omega_{w}$ and hence of $\omega_{110}$ and $c_{110}$.


If the biquadratic patch is in fact part of a quadric surface, 
Theorem~1 provides its bilinear forms and we can calculate its 
geometric elements. We see it with an example:

\begin{figure}
\centering
\includegraphics[height=5cm]{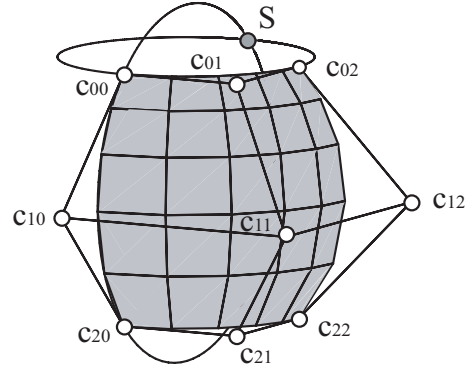}
\caption{Spherical rational tensor product patch}
\label{esfera4}
\end{figure}
\begin{example} Net:%
$\left[\begin{array}{ccc} (a , 0 ,a) &
(a , a , a) &( 0, a, a)\\
( 2a, 0, 0)& (2a, 2a, 0) & ( 0, 2a, 0)\\
(a , 0 ,-a) &
(a, a , -a) &( 0, a, -a)
\end{array}\right]$ with $a=1/\sqrt{2}$ and  
 weights: 
 $\left[\begin{array}{ccc}1 & 1 & 2 \\1/2 & 1/2 & 1\\ 1/2 & 1/2 &1
 \end{array}\right]$ (Fig.~\ref{esfera4}):\end{example}
 
We use a triangular patch through $P=c_{00}$, $Q=c_{02}$, $R=c_{20}$ with 
the following control net and weights
\[\left[\begin{array}{c}(a,0,a)\hspace{0.1cm}(a,a,a)
\hspace{0.1cm}(0,a,a)\\(2a,0,0)\hspace{0.1cm}c_{110}\\ (a,0,-a)\end{array}\right]  
,\quad 
\left[\begin{array}{c}1\hspace{0.5cm}1\hspace{0.5cm}2\\1/2\hspace{0.1cm}\omega_{110}\\ 
1/2\end{array}\right],\]
and notice that the conic at $u$ and the conic at $v$ meet at the 
point $S=(-\sqrt{2}/2, 0, \sqrt{2}/2)$ and
\begin{eqnarray*}\hspace{-8mm}S=
c_{00}-2c_{01}+2c_{02} =
\frac{c_{00}-c_{10}+c_{20}/2}{1/2}.
\end{eqnarray*}

There is no need to perform M\"obius transformations, since the weights 
already satisfy
\begin{eqnarray*}\hspace{-7mm}
\omega_{u}=\omega_{00}-2\omega_{01}+ \omega_{02}=1,\quad 
\omega_{v}=\omega_{00}-2\omega_{10}+ \omega_{20}=\frac{1}{2},
\end{eqnarray*}
but the other denominator is not determined,
\[\omega_{w}=\omega_{20}-2\omega_{110}+\omega_{02}=\frac{5}{2}-2\omega_{110}.\]

The control points and $S$ define the planes $p,q,r,t$ and their 
intersections,
\[U= \left(\frac{\sqrt{2}}{4}, \frac{\sqrt{2}}{4}, 
\frac{3\sqrt{2}}{4}\right), \quad 
V=\left(\sqrt{2},-\frac{\sqrt{2}}{2},0\right),\]
\[ T=(1,\sqrt{2},\sqrt{2},0),\]
except for $W$, which is a point at infinity with direction $(1,-1,1)$.

This means that $\Omega_{W}=1$,
but the normalisation term for $W$
\[-\omega_{002}\omega_{w}+\omega_{020}\omega_{v}+\omega_{200}\omega_{u}=
2\omega_{110}-1\] must vanish and hence $\omega_{110}=1/2$, 
$\omega_{w}=3/2$ and the representative for $W$ is the vector
\[W=-\omega_{00}\omega_{w}P+\omega_{02}\omega_{v}Q+\omega_{21}\omega_{u}=
\frac{\sqrt{2}}{2}\left(-1,1,-1\right).\]

The coefficients
\[\Omega_{U}=2,\quad \Omega_{V}=1,\quad \Omega_{W}=1,\quad 
\lambda=2,\] yield the expression for the bilinear form for the 
quadric,
\[\mathcal Q=p^2+q^2+\frac{r^2}{4}-2pq-pr-qr+2 t^2.\]

 The normalised forms for the planes are
\[p=-\frac{\sqrt{2}}{2}(x+z-\sqrt{2}),\quad 
q=\frac{\sqrt{2}}{3}(\sqrt{2}-y-z),\]
\[r=\frac{\sqrt{2}}{3}(\sqrt{2}-x+z),\quad 
t=\frac{\sqrt{2}}{3}\left(x+y-\frac{\sqrt{2}}{2}\right),\]
and so the implicit equation for the surface in cartesian coordinates 
is \[x^2+y^2+z^2 -1=0.\]

\section{Conclusions}

We have derived closed formulas in terms of control points and weights
for several geometric elements of quadrics in rational B\'ezier form,
both in triangular and tensor product representation.  To our
knowledge, these formulas have not been produced before.  The main
difference with other procedures for drawing geometric information
from rational triangular patches \cite{sederberg} is the use of
geometric entities such as tangent planes to the quadric and their
intersections as ingredients for obtaining bilinear forms, and hence,
implicit equations, for the surface.  There are many ways of
implicitising a parametric surface \cite{sederberg-implicit}, but the use of linear forms with
geometrical meaning instead of cartesian coordinates simplifies this
problem for quadric patches.  Besides, these geometric entities appear
naturally in the formulas for geometric elements because they are
already present in the expressions for the bilinear forms for the
quadric.  The use of projective algebraic geometry allows us to
perform calculations in a synthetic fashion, instead of resorting to
cartesian coordinates.

Additionally we classify affine quadrics using one coefficient involving
the weights of the patch. This can be done without implicitising the 
quadric patch \cite{gudrunquadric, gudruninvariants}, but the closed 
form for the implicit equations is what enables us to derive closed formulas 
for geometric elements.

The results are obtained initially for B\'ezier triangles, but are
also extended to quadric patches in tensor product form.

\bibliographystyle{elsarticle-num-names}
\bibliography{cagd}

\begin{thebibliography}{23}
\providecommand{\natexlab}[1]{#1}
\providecommand{\url}[1]{\texttt{#1}}
\providecommand{\urlprefix}{URL }
\expandafter\ifx\csname urlstyle\endcsname\relax
  \providecommand{\doi}[1]{doi:\discretionary{}{}{}#1}\else
  \providecommand{\doi}[1]{doi:\discretionary{}{}{}\begingroup
  \urlstyle{rm}\url{#1}\endgroup}\fi
\providecommand{\bibinfo}[2]{#2}

\bibitem[{Pottmann et~al.(2007)Pottmann, Asperl, Hofer, and
  Kilian.}]{architecture}
\bibinfo{author}{H.~Pottmann}, \bibinfo{author}{A.~Asperl},
  \bibinfo{author}{M.~Hofer}, \bibinfo{author}{A.~Kilian.},
  \bibinfo{title}{Architectural geometry}, \bibinfo{publisher}{Bentley
  Institute Press}, \bibinfo{address}{Exton}, \bibinfo{year}{2007}.

\bibitem[{Farin(1986)}]{triangle}
\bibinfo{author}{G.~Farin}, \bibinfo{title}{Triangular {B}ernstein-{B}\'ezier
  patches.}, \bibinfo{journal}{Comput. Aided Geom. Design}
  \bibinfo{volume}{3}~(\bibinfo{number}{2}) (\bibinfo{year}{1986})
  \bibinfo{pages}{83--127}, ISSN \bibinfo{issn}{0167-8396}.

\bibitem[{Sederberg and Anderson(1985)}]{sederberg}
\bibinfo{author}{T.~Sederberg}, \bibinfo{author}{D.~Anderson},
  \bibinfo{title}{{S}teiner Surface Patches}, \bibinfo{journal}{IEEE Computer
  Graphics and Applications} \bibinfo{volume}{5} (\bibinfo{year}{1985})
  \bibinfo{pages}{23--36}.

\bibitem[{Boehm and Hansford(1991)}]{hansford}
\bibinfo{author}{W.~Boehm}, \bibinfo{author}{D.~Hansford},
  \bibinfo{title}{{B}\'ezier Patches on Quadrics}, in:
  \bibinfo{editor}{G.~Farin} (Ed.), \bibinfo{booktitle}{{NURBS} for Curves and
  Surface Design}, \bibinfo{publisher}{SIAM}, \bibinfo{pages}{1--14},
  \bibinfo{year}{1991}.

\bibitem[{Lodha and Warren(1990)}]{lodha}
\bibinfo{author}{S.~Lodha}, \bibinfo{author}{J.~Warren},
  \bibinfo{title}{{B}\'ezier representation for quadric surface patches},
  \bibinfo{journal}{Computer-Aided Design}
  \bibinfo{volume}{22}~(\bibinfo{number}{9}) (\bibinfo{year}{1990})
  \bibinfo{pages}{574 -- 579}.

\bibitem[{Dietz et~al.(1993)Dietz, Hoschek, and J\"uttler}]{dietz}
\bibinfo{author}{R.~Dietz}, \bibinfo{author}{J.~Hoschek},
  \bibinfo{author}{B.~J\"uttler}, \bibinfo{title}{An algebraic approach to
  curves and surfaces on the sphere and on other quadrics},
  \bibinfo{journal}{Computer Aided Geometric Design}
  \bibinfo{volume}{10}~(\bibinfo{number}{3-4}) (\bibinfo{year}{1993})
  \bibinfo{pages}{211 -- 229}.

\bibitem[{Dietz et~al.(1995)Dietz, Hoschek, and J\"uttler}]{dietz1}
\bibinfo{author}{R.~Dietz}, \bibinfo{author}{J.~Hoschek},
  \bibinfo{author}{B.~J\"uttler}, \bibinfo{title}{Rational patches on quadric
  surfaces}, \bibinfo{journal}{Computer-Aided Design}
  \bibinfo{volume}{27}~(\bibinfo{number}{1}) (\bibinfo{year}{1995})
  \bibinfo{pages}{27 -- 40}.

\bibitem[{Coffman et~al.(1996)Coffman, Schwartz, and Stanton}]{coffman}
\bibinfo{author}{A.~Coffman}, \bibinfo{author}{A.~J. Schwartz},
  \bibinfo{author}{C.~Stanton}, \bibinfo{title}{The algebra and geometry of
  {S}teiner and other quadratically parametrizable surfaces},
  \bibinfo{journal}{Computer Aided Geometric Design}
  \bibinfo{volume}{13}~(\bibinfo{number}{3}) (\bibinfo{year}{1996})
  \bibinfo{pages}{257 -- 286}.

\bibitem[{Degen(1996)}]{degen}
\bibinfo{author}{W.~Degen}, \bibinfo{title}{The Types of Triangular {B}\'ezier
  Surfaces}, in: \bibinfo{editor}{G.~Mullineux} (Ed.), \bibinfo{booktitle}{The
  Mathematics of Surfaces {VI}}, \bibinfo{publisher}{Clarendon Press},
  \bibinfo{pages}{153--170}, \bibinfo{year}{1996}.

\bibitem[{Albrecht(1998{\natexlab{a}})}]{gudrunquadric}
\bibinfo{author}{G.~Albrecht}, \bibinfo{title}{Determination and classification
  of triangular quadric patches}, \bibinfo{journal}{Computer Aided Geometric
  Design} \bibinfo{volume}{15}~(\bibinfo{number}{7})
  (\bibinfo{year}{1998}{\natexlab{a}}) \bibinfo{pages}{675 -- 697}.

\bibitem[{Albrecht(1998{\natexlab{b}})}]{gudrunIEEE}
\bibinfo{author}{G.~Albrecht}, \bibinfo{title}{Rational quadratic Bezier
  triangles on quadrics}, in: \bibinfo{editor}{F.-E. Wolter},
  \bibinfo{editor}{N.~Patrikalakis} (Eds.), \bibinfo{booktitle}{Computer
  Graphics International, 1998. Proceedings}, \bibinfo{publisher}{IEEE},
  \bibinfo{address}{Los Alamitos, CA}, \bibinfo{pages}{34--40},
  \bibinfo{year}{1998}{\natexlab{b}}.

\bibitem[{Albrecht(2011)}]{gudruninvariants}
\bibinfo{author}{G.~Albrecht}, \bibinfo{title}{Geometric invariants of
  parametric triangular quadric patches}, \bibinfo{journal}{International
  Electronic Journal of Geometry} \bibinfo{volume}{4}~(\bibinfo{number}{2})
  (\bibinfo{year}{2011}) \bibinfo{pages}{63 -- 84}.

\bibitem[{S\'anchez-Reyes and Paluszny(2000)}]{reyes-quadrics}
\bibinfo{author}{J.~S\'anchez-Reyes}, \bibinfo{author}{M.~Paluszny},
  \bibinfo{title}{Weighted radial displacement: A geometric look at {B}\'ezier
  conics and quadrics}, \bibinfo{journal}{Computer Aided Geometric Design}
  \bibinfo{volume}{17}~(\bibinfo{number}{3}) (\bibinfo{year}{2000})
  \bibinfo{pages}{267 -- 289}.

\bibitem[{Albrecht(2004)}]{gudrun-parametric}
\bibinfo{author}{G.~Albrecht}, \bibinfo{title}{An Algorithm for Parametric
  Quadric Patch Construction}, \bibinfo{journal}{Computing}
  \bibinfo{volume}{72}~(\bibinfo{number}{1-2}) (\bibinfo{year}{2004})
  \bibinfo{pages}{1--12}.

\bibitem[{Albrecht et~al.(2015)Albrecht, Paluszny, and Lentini}]{gudrunmarco}
\bibinfo{author}{G.~Albrecht}, \bibinfo{author}{M.~Paluszny},
  \bibinfo{author}{M.~Lentini}, \bibinfo{title}{An intuitive way for
  constructing parametric quadric triangles}, \bibinfo{journal}{Computational
  and Applied Mathematics}  (\bibinfo{year}{2015}) \bibinfo{pages}{1--23}.

\bibitem[{Cant\'on et~al.(2011)Cant\'on, Fern\'andez-Jambrina, and
  Rosado-Mar{\'\i}a}]{conics}
\bibinfo{author}{A.~Cant\'on}, \bibinfo{author}{L.~Fern\'andez-Jambrina},
  \bibinfo{author}{E.~Rosado-Mar{\'\i}a}, \bibinfo{title}{Geometric
  characteristics of conics in {B}\'ezier form},
  \bibinfo{journal}{Computer-Aided Design}
  \bibinfo{volume}{43}~(\bibinfo{number}{11}) (\bibinfo{year}{2011})
  \bibinfo{pages}{1413 -- 1421}.

\bibitem[{Albrecht(1999)}]{gudruntesis}
\bibinfo{author}{G.~Albrecht}, \bibinfo{title}{Rational Triangular B\'ezier
  Surfaces - Theory and Applications}, \bibinfo{type}{Habilitationschrift},
  \bibinfo{school}{Fakult\"at f\"ur Mathematik, TU M\"unchen, Shaker-Verlag,
  Aachen}, \bibinfo{year}{1999}.

\bibitem[{Farin(2002)}]{farin}
\bibinfo{author}{G.~Farin}, \bibinfo{title}{Curves and surfaces for CAGD: a
  practical guide}, \bibinfo{publisher}{Morgan Kaufmann Publishers Inc.},
  \bibinfo{address}{San Francisco, CA, USA}, \bibinfo{edition}{5th} edn., ISBN
  \bibinfo{isbn}{1-55860-737-4}, \bibinfo{year}{2002}.

\bibitem[{Cant\'on et~al.(2015)Cant\'on, Fern\'andez-Jambrina,
  Rosado~Mar\'{\i}a, and V\'aquez-Gallo}]{oldquadric}
\bibinfo{author}{A.~Cant\'on}, \bibinfo{author}{L.~Fern\'andez-Jambrina},
  \bibinfo{author}{E.~Rosado~Mar\'{\i}a}, \bibinfo{author}{M.~V\'aquez-Gallo},
  \bibinfo{title}{Implicit Equations of Non-degenerate Rational B\'ezier
  Quadric Triangles}, in: \bibinfo{editor}{J.-D. Boissonnat},
  \bibinfo{editor}{A.~Cohen}, \bibinfo{editor}{O.~Gibaru},
  \bibinfo{editor}{C.~Gout}, \bibinfo{editor}{T.~Lyche}, \bibinfo{editor}{M.-L.
  Mazure}, \bibinfo{editor}{L.~L. Schumaker} (Eds.), \bibinfo{booktitle}{Curves
  and Surfaces}, vol. \bibinfo{volume}{9213} of \emph{\bibinfo{series}{Lecture
  Notes in Computer Science}}, \bibinfo{publisher}{Springer International
  Publishing}, ISBN \bibinfo{isbn}{978-3-319-22803-7}, \bibinfo{pages}{70--79},
  \bibinfo{year}{2015}.

\bibitem[{Semple and Kneebone(1952)}]{kneebone}
\bibinfo{author}{J.~G. Semple}, \bibinfo{author}{G.~T. Kneebone},
  \bibinfo{title}{Algebraic projective geometry}, \bibinfo{publisher}{Oxford
  University Press}, \bibinfo{address}{London}, \bibinfo{year}{1952}.

\bibitem[{Boehm(1993)}]{boehm-quadrics}
\bibinfo{author}{W.~Boehm}, \bibinfo{title}{Some remarks on quadrics},
  \bibinfo{journal}{Computer Aided Geometric Design}
  \bibinfo{volume}{10}~(\bibinfo{number}{3--4}) (\bibinfo{year}{1993})
  \bibinfo{pages}{231 -- 236}.

\bibitem[{Fink(1992)}]{fink}
\bibinfo{author}{U.~Fink}, \bibinfo{title}{Biquadratische
  B\'ezier-Fl\"achenst\"ucke auf Quadriken}, Master's thesis,
  \bibinfo{school}{Fakult\"at Mathematik der Universit\"at Stuttgart},
  \bibinfo{address}{Mathematisches Institut der Universit\"at Stuttgart},
  \bibinfo{year}{1992}.

\bibitem[{Sederberg et~al.(1984)Sederberg, Anderson, and
  Goldman}]{sederberg-implicit}
\bibinfo{author}{T.~Sederberg}, \bibinfo{author}{D.~Anderson},
  \bibinfo{author}{R.~Goldman}, \bibinfo{title}{Implicit representation of
  parametric curves and surfaces}, \bibinfo{journal}{Computer Vision, Graphics,
  and Image Processing} \bibinfo{volume}{28}~(\bibinfo{number}{1})
  (\bibinfo{year}{1984}) \bibinfo{pages}{72 -- 84}.

\end{thebibliography}

\end{document}